\documentclass[fleqn,usenatbib]{mnras}

\usepackage{newtxtext,newtxmath}

\usepackage[T1]{fontenc}
\usepackage{ae,aecompl}

\usepackage{graphicx}	
\usepackage{amsmath}	
\usepackage[caption=false]{subfig}

\title[\texttt{ARIADNE}]{\texttt{ARIADNE}: Measuring accurate and precise stellar parameters through SED fitting}

\author[J. I. Vines.]{
Jose I. Vines$^{1}$\thanks{E-mail: jose.vines@ug.uchile.cl}
\& James S. Jenkins$^{1,2,3}$
\\
$^1$Departamento de Astronom\'ia, Universidad de Chile, Casilla 36-D, Santiago, Chile\\
$^2$N\'ucleo de Astronom\'ia, Facultad de Ingenier\'ia y Ciencias, Universidad Diego Portales, Av. Ej\'ercito 441, Santiago, Chile\\
$^3$Centro de Astrof\'isica y Tecnolog\'ias Afines (CATA), Casilla 36-D, Santiago, Chile\\
}

\date{Accepted XXX. Received YYY; in original form ZZZ}

\pubyear{2015}
\hypersetup{draft}
\begin{document}
\label{firstpage}
\pagerange{\pageref{firstpage}--\pageref{lastpage}}
\maketitle

\begin{abstract}

Accurately measuring stellar parameters is a key goal to increase our understanding of the observable universe. However, current methods are limited by many factors, in particular, the biases and physical assumptions that are the basis for the underlying evolutionary or atmospheric models, those that these methods rely upon. Here we introduce our code spectrAl eneRgy dIstribution bAyesian moDel averagiNg fittEr (\texttt{ARIADNE}), which tackles this problem by using Bayesian Model Averaging to incorporate the information from all stellar models to arrive at accurate and precise values. This code uses spectral energy distribution fitting methods, combined with precise Gaia distances, to measure the temperature, $\log{\rm g}$, [Fe/H], A$_{\rm V}$, and radius of a star. When compared with interferometrically measured radii \texttt{ARIADNE} produces values in excellent agreement across a wide range of stellar parameters, with a mean fractional difference of only 0.001~$\pm$~0.070. We currently incorporate six different models, and in some cases we find significant offsets between them, reaching differences of up to 550~K and 0.6~R$_\odot$ in temperature and radius, respectively.  For example, such offsets in stellar radius would give rise to a difference in planetary radius of 60\%, negating homogeneity when combining results from different models. We also find a trend for stars smaller than 0.4-0.5~R$_\odot$, which shows more work needs to be done to better model these stars, even though the overall extent is within the uncertainties of the interferometric measurements. We advocate for the use of \texttt{ARIADNE} to provide improved bulk parameters of nearby A to M dwarfs for future studies.

\end{abstract}

\begin{keywords}
stars:atmospheres -- methods:data analysis -- stars:fundamental parameters
\end{keywords}



\section{Introduction}

Stellar characterization has historically been an important and difficult task in astronomy, and tremendous effort has been made to precisely constrain stellar bulk parameters. Such efforts include space-based missions to measure the distance to stars through their parallax \citep{hipparcos, VanLeeuwen2007, GAIA}, directly observing the stellar angular diameter through interferometry in order to obtain precise radii and effective temperatures \citep{Lane2001, Berger2006, Boyajian2012, Boyajian2012b, Boyajian2013, Boyajian2015, Tanner2015, Ligi2016, Bonnefoy2018, Ligi2019}, analysing high-precision space-based photometry for asteroseismic signals to determine radius and mass \citep{Huber2013, Hernandez2016, Campante2019, Stokholm2019}, and analysing high-resolution spectra to constrain the metallicity and other chemical abundances, temperature, and rotation velocities through equivalent width calculations, spectral synthesis, wavelet analysis, or other techniques \citep[e.g.][]{1992oasp.book.....G, 1993A&A...275..101E, Valenti1996, 2005ApJS..159..141V, Jenkins2008, Pavlenko2012, Yee2017, Soto2018, Gill2018, 2019A&A...628A.131T}.
Some of these methods are subject to strict restrictions; for example interferometry is only feasible for the nearest bright stars, with similar biases related to parallax measurements, whereas spectral analysis requires high signal-to-noise ratio spectra, which is not always available, or requires large blocks of observing time to acquire. 

Another method used to determine physical stellar parameters is by comparing the shape of the spectral energy distribution (SED), constructed from different photometric observations, to those of synthetic stellar atmospheric models (for examples, see: \citealt{Robitaille2007, Pecaut2013, NGTS_1b, Morrell2019, Martinez-Rodriguez2019}). There have been several different approaches to fitting stellar SEDs, for example \cite{Masana2006} introduced Spectral Energy Distribution Fit (SEDF) where they fit for the SED of stars using VJHK photometry by minimizing a specific $\chi^2$ function defined by them. Another $\chi^2$ minimization approach is the Virtual Observatory SED Analyzer (VOSA; from the Spanish Virtual Observatory; \citealt{Bayo2008}), a tool that allows users to fit for different atmospheric models online. \cite{Stassun2016} also present a $\chi^2$ approach, but instead of fitting for all parameters, they instead fix T$_{\rm eff}$, $\log{\rm g}$ and [Fe/H] and fit only for the extinction A$_{\rm V}$ and overall normalization, thus claiming virtually model-independent parameters \citep{Stassun2016b, Stevens2017, Stassun2018}. A Markov Chain Monte Carlo (MCMC) approach to fitting SEDs has been implemented by \cite{Gillen2017} where they fit broad-band photometry to PHOENIX v2 \citep{Husser2013} and BT-Settl \citep{Allard2012} models. 

\citet{Torres2010} highlights the scatter and offsets produced by uncertainties in evolutionary models and chemical abundances (see their Figures 6 and 7) by studying detached binary systems with accurate mass and radii estimations, and developed empirical relations between radius/mass and other stellar parameters to a precision of around $6\%$. While this provides precise masses and radii, it requires rather precise knowledge of T$_{\rm eff}$, $\log{\rm g}$, and [Fe/H], which in turn can be affected by biases in stellar modeling as well, in particular for giant stars \citep{Lebzelter2012}. Some aspects of stellar modeling that could introduce biases are the fact that models such as \cite{1993KurCD..13.....K} and \cite{Castelli2004} have tabulated opacities which are interpolated during the atmosphere modeling, whereas models such as the ones computed from the PHOENIX code \citep{Allard2012, Husser2013} calculate the opacities on the fly. In addition, these models also assume different stellar abundances.  For example, the BT-Settl and BT-NextGen models both use stellar abundances from \citet{ASPLUND09}, while the BT-Cond models use abundances from \citet{Caffau11}. There are also differences in the assumed microphysics in each model, such as the assumption of LTE or NLTE, the addition of convection cells and/or overshooting, and the employed geometries. Furthermore \citet{Basu2012} studied the effects of the different input physics of several stellar models on the estimation of mass and radius and found significant offsets in these parameters when they were estimated from stellar models alone.  They conclude that an accuracy of just $8\%$ and $14\%$ can be attained for mass and radius respectively, in the best case, even assuming perfect knowledge of the stellar micro-physics. Since planetary radii derived from transits scale linearly with the stellar radii, and mass scales as $M^{2/3}$, this means that there is a potential offset of $14\%$ for planetary radii and $5\%$ for planetary mass. However, to the best of our knowledge there has not been studies comparing different stellar atmosphere model grids.

In the era of high-precision radial-velocity and photometric exoplanet search programs, such as Coralie \citep{CORALIE}, HARPS \citep{Mayor2003}, CHEPS \citep{Jenkins2009}, ESPRESSO \citep{Pepe2013}, WASP \citep{Pollacco2006}, NGTS \citep{NGTS_2018}, \textit{Kepler} \citep{KEPLER}, \textit{K2} \citep{Howell2014}, TESS \citep{TESS} and the upcoming PLATO mission \citep{PLATO}, precise knowledge of the stellar parameters has taken on a hugely significant role in the characterization of any detected systems, due to planets having their bulk parameters defined in terms of the host star's. Studying stellar parameters does not only shed light on any planetary companions bulk parameters, but they also allow us to study their composition \citep{Zeng2008, Lopez2014, Thorngren2016, Zeng2016, Bashi2017, Otegi2019, Thorngren2019}, formation and evolutionary models \citep{Ida2004, Fischer2005, Mekias2008, Buchhave2012, Jackson2012, Jenkins2013, Valsecchi2014, Arras2017, Schoonenberg2019}, and even future evolution of the system \citep{Essick2015, Burdge2019}.

In this work, we aim to address the problem of systematic biases that are introduced by employing different models as part of our Bayesian Model Averaging (BMA) approach that we have implemented in our package \texttt{ARIADNE}\footnote{https://github.com/jvines/astroARIADNE}.  We describe our method to include the information from all models when calculating stellar parameters, maximising our accuracy and precision. This paper has the following structure: In Section \ref{sec:convolution} we describe the models, data used (Sect. \ref{sec:photometry_retrieval}, prior selection and sampling algorithm (Sects. \ref{sec:priors} and \ref{sec:sampling} respectively), the BMA framework (Sect. \ref{sec:bma}) and the main driver algorithm behind \texttt{ARIADNE} in Section \ref{sec:algorithm}; in Section \ref{sec:benchmark} we show the comparison stars used to validate the code and in Section \ref{sec:results} we discuss the fundamental stellar properties: radius, effective temperature, $\log{\rm g}$, [Fe/H] and mass; in Section \ref{sec:offset} we highlight the models systematic offsets and in Section \ref{sec:conclusion} we summarize and highlight future work.

\section{SED Modelling with \texttt{ARIADNE}}

\label{sec:convolution}

\texttt{ARIADNE} is designed to automatically fit the SEDs of stars using up to six different atmospheric model grids, in order to obtain effective temperature, surface gravity, metallicity, distance, radius and V-band extinction. We also provide computed parameters such as stellar luminosity and gravitational mass, and isochrone interpolated age and mass. The only required inputs for \texttt{ARIADNE} are the coordinates of the object and optionally, the Gaia DR2 id.

To create the model grids we convolved PHOENIX v2 \citep{Husser2013}, BT-Settl \citep{Allard2012}, BT-NextGen \citep{Hauschildt1999,Allard2012}, BT-Cond \citep{Allard2012}, Castelli \& Kurucz \citep{Castelli2004}, and Kurucz \citep{1993KurCD..13.....K} stellar atmosphere models with commonly available broadband photometry bandpasses: 
\begin{itemize}
    \item Johnson UBV
    \item Tycho-2 BtVt
    \item Cousins RI
    \item 2MASS JHK$_\text{s}$
    \item SDSS \textit{ugriz}
    \item ALL-WISE W1 and W2
    \item Gaia DR2v2 G, RP and BP
    \item Pan-STARRS1 \textit{griwyz}
    \item GALEX NUV and FUV
    \item Spitzer/IRAC 3.6$\mu$m and 4.5$\mu$m
    \item Str\"omgren uvby
    \item TESS
    \item Kepler
    \item NGTS
\end{itemize}

Each SED is modeled with $6 + n$ parameters, highlighted in Equation \ref{eq:model}, written in standard statistical notation, where $n$ is the number of available photometry measurements. S is the model grid which we interpolate in T$_\text{eff}$ -- $\log{\rm g}$ -- [Fe/H] space, and $\epsilon_i$ is the noise model of each $i$th photometry measurement that accounts for possible photometric uncertainty underestimations (due to variability, for example), which we expand in Equation \ref{eq:excess}, where $\sigma_{m,i}$ is the measurement error of the $i$th photometry observation, and $\sigma_{e,i}$ is the excess noise of the corresponding observation. The fluxes in each grid are given in FLAM units (erg cm$^{-2}$ s$^{-2}$ $\mu$m$^{-1}$) at the surface of the star, so to obtain observed fluxes the SED is multiplied by a normalization factor given by $(R/D)^2$ where $R$ is the radius of the star and $D$ is the distance to the star.  We also offer the possibility of fitting for the normalization factor directly instead of radius and distance, and calculate the radius afterwards using the parallax if available. We also model the interstellar extinction A$_\text{V}$ by offering four different extinction laws: \cite{ccm89, odonnell94, calzetti01, fitzpatrick99} provided by the extinction package\footnote{https://extinction.readthedocs.io/en/latest/}, all of them with fixed ratio of total to selective absorption R$_\text{V} = 3.1$. We chose to fix the extinction factor due to it's negligible effect in the overall shape of the SED, though we leave open the possibility of inclusion as an extra parameter in future versions.

\begin{equation}
\label{eq:model}
    SF = 10^{-0.4{\rm A}_{\rm i}} \cdot S(T_{\rm eff}, \log{\rm g}, {\rm [Fe/H]}) \cdot \left(\frac{R}{D}\right)^2 + \epsilon_i
\end{equation}

\begin{equation}
\label{eq:excess}
    \epsilon_i^2 = \sigma_{m,i}^2 + \sigma_{e,i}^2
\end{equation}

Table \ref{tab:limits} shows the limits in T$_{\rm eff}$, $\log{\rm g}$ and [Fe/H] for each model. Even though we have constrained the upper limit for T$_{\rm eff}$ for all models to a value of 12,000 K for homogeneity, BT-models and the Kurucz models could be extended to higher temperatures in a future update.

\begin{table}
\caption{SED model limits.}              
\label{tab:limits}      
\centering   
\begin{tabular}{c c c c}          
\hline\hline                        
Model & T$_{\rm eff}$ & $\log{\rm g}$ & [Fe/H] \\
\hline 
Phoenix v2  & [2300, 12000] & [0.0, 6.0] & [-2.0, 1.0] \\
BT-Settl & [2300, 12000] & [-0.5, 6.0] & [-1.0, 0.5] \\
BT-NextGen & [2300, 12000] & [-0.5, 6.0] & [-1.0, 0.5] \\
BT-Cond & [2300, 12000] & [-0.5, 6.0] & [-2.5, 0.5] \\
Castelli \& Kurucz & [3500, 12000] & [0.0, 5.0] & [-2.5, 0.5] \\
Kurucz & [3500, 12000] & [0.0, 5.0] & [-1.0, 1.0] \\

\hline \\
\end{tabular}
\end{table}

\subsection{Automatic photometry retrieval}
\label{sec:photometry_retrieval}

The code searches automatically for available photometry using \texttt{astroquery}\footnote{https://astroquery.readthedocs.io/en/latest/} to access the Gaia DR2 (\citealt{GAIA, GAIA_DR2}) archive (for parallax, T$_\text{eff}$, and radius when available), MAST and VizieR \footnote{http://vizier.u-strasbg.fr/} to query catalogs APASS DR9 \citep{APASS}, ASCC \citep{ASCC}, ALL-WISE \citep{WISE}, 2MASS \citep{2MASS}, SDSS DR12 \citep{SDSS}, Pan-STARRS1 \citep{PS1}, GALEX \citep{GALEX}, Str\"omgren Photometric Catalog \citep{Paunzen15}, GLIMPSE \citep{GLIMPSE}, Tycho-2 \citep{2000A&A...355L..27H} and Gaia DR2 for the magnitudes detailed in Section \ref{sec:convolution}. In addition to recovering the magnitudes, we also perform a quality check and reject sources flagged as extended or bad quality by the quality filters of their respective catalogs. In the case where there are magnitudes with no reported uncertainty, we assume they are highly uncertain and calculate their flux uncertainty as $\delta F_i = 1.1 F_i \max(\delta F / F)$, where $F_i$ is the $i$th flux and $F$ refers to all of the fluxes in the data set. To ensure that we are recovering the correct photometry for the star, we turn to several catalogs that we crossmatch with Gaia, including APASS, ASCC, ALL-WISE, 2MASS, SDSS and Pan-STARRS1. The ASCC crossmatch provides a Tycho-2 identifier which is used in its homonym's and Str\"omgren catalogs, the GLIMPSE catalog can be searched with the 2MASS identifier recovered from the Gaia-2MASS crossmatch catalog, and finally in the case of GALEX, we do the crossmatch with Gaia using the xMatch query module found in \texttt{astroquery}, which serves as an API for the CDS X-Match service.

\subsection{Prior selection}
\label{sec:priors}

 \texttt{ARIADNE} offers flexibility to the user to choose priors for each parameter, but also provides default priors. For the effective temperature, $\log{\rm g}$, and [Fe/H] we use empirical priors derived from their respective distributions observed by the fifth RAVE survey data release \citep{RAVE}.  In the case of the $\log{\rm g}$ the possibility of estimating the value using MESA Isochrones \& Stellar Tracks isochrones with the \texttt{isochrones}\footnote{https://isochrones.readthedocs.io/en/latest/} package (MIST; \citealt{isochrones, MIST}) exists, where here \texttt{ARIADNE} makes use of the Gaia estimates for effective temperature, luminosity and parallax when available, and a subset of the photometry to carry out the interpolations. For the distance we use the Gaia distance derived by \cite{Bailer-Jones2018} (hereafter BJ18) to constrain a normally distributed prior using five times the largest uncertainty as the standard deviation in order to account for offsets in the distance, to avoid reaching machine precision and thus obtaining unrealistically precise posteriors, and for the radius we employ an uninformative flat prior ranging from 0.05 to 100 R$_\odot$. The extinction is limited to the maximum line-of-sight extinction from the updated SFD galactic dust map (\citealt{Schlegel1998, Schlafly2011}) through the \texttt{dustmaps} python package \citep{2018JOSS....3..695M}. The excess noise terms have zero mean gaussian priors, with three times the squared uncertainty set as the variance, and finally the normalization factor has an uninformative prior. We show a quick summary of the priors in Table \ref{tab:priors}.

Though we offer up to six atmospheric model grids, there are restrictions on which ones can be used. \cite{1993KurCD..13.....K} and \cite{Castelli2004} models are known to be unreliable with low temperature stars, thus they can't be chosen as models for stars with temperatures below 4000 K. On the other hand, the BT-suite of models don't show significant variation for temperatures above 4000 K (Figures 3 and 4 from \citealt{Allard2012}), meaning that applying the three models on stars in this temperature regime can induce bias in the final stellar parameter uncertainties, which led to our decision of limiting the BT-suite of models to only BT-Settl for these types of stars.

\begin{table}
\caption{\texttt{ARIADNE} default priors}
\centering
\label{tab:priors}
\begin{tabular}{ll}
\hline
Parameter                         & Prior                   \\ \hline
T$_{\rm eff}^\dagger$ & RAVE \\
$\log$\,g & $\mathcal{N}(g, g_e^2)$ \\
$\rm{[Fe/H]}$ & $\mathcal{N}(-0.125, 0.234^2)$ \\
$D$ & $\mathcal{N}(D, (5D_e)^2)$ \\
$R_*$ & $\mathcal{U}(0.05, 100)$ \\
A$_{\rm V}$ & $\mathcal{U}(0, av)$ \\ \hline
\hline
\end{tabular}\\
Where $g,\,g_e$ is the MIST $\log\,$g estimate and its uncertainty, $D,\,D_e$ is the BJ18 distance estimate with the largest errorbar, and $av$ is the largest line-of-sight extinction in the SFD galactic dust map. For details see the text. \\$\dagger$ The effective temperature prior is drawn from the empirical temperature distribution from the fifth RAVE data release.
\end{table}

\subsection{Bayesian Model Averaging}
\label{sec:bma}

BMA is an extension to the usual Bayesian inference framework, where one does not only fit a single model to obtain parameters and uncertainties (and sometimes the so-called model evidence), but one also fits different models and can potentially account for model-specific systematic biases, allowing for model selection, prediction and combined estimation; the latter being the focus in this paper \citep{BMA}. Firstly, recall Bayes' theorem:

\begin{equation}
\label{eq:bma_1}
    P\left(\hat{\theta} | X, M\right) = \frac{P\left(X | \hat{\theta}, M\right) P\left(\hat{\theta} | M\right)}{P(X | M)}
\end{equation}

Where $X$ is the observed data, $\hat{\theta}$ is the parameter vector corresponding to model $M$, $P\left(X | \hat{\theta}, M\right)$ is the likelihood function, $P\left(\hat{\theta} | M\right)$ is the prior function, and the denominator $P(X | M) = \int P\left(X | \hat{\theta}, M\right) P\left(\hat{\theta} | M\right)d\hat{\theta}$ is the marginalized likelihood of model $M$ with parameters $\hat{\theta}$, which is also referred to as the model evidence or Bayesian evidence; this quantity lies at the heart of BMA. It is possible to derive posterior probabilities for different models $M_j$ by applying Bayes' theorem as shown in Equation \ref{eq:bma_2}. The combined distribution for the $j$th parameter, $\Bar{\theta}_j$, is then given by the weighted average of the $N$ considered models, where the weight of the $n$th model is its posterior probability, as shown in Equation \ref{eq:bma_3}. In this work we assume that all models are equally probable, a priori, so Equation \ref{eq:bma_2} reduces to normalizing the evidence of each model by the sum of all the available evidences.

\begin{equation}
\label{eq:bma_2}
    P(M_j | X) = \frac{P(X | M_j) P (M_j)}{\sum^N_{n=1} P(X | M_n) P(M_n)}
\end{equation}

\begin{equation}
\label{eq:bma_3}
    P(\Bar{\theta}) = \sum^N_{n=1}P(\hat{\theta}_n | X, M_n) P(M_n | X)
\end{equation}

For \texttt{ARIADNE}, the likelihood function is gaussian, as defined in Equation \ref{eq:likelihood}, where $f_i$ is the $i$th measured flux, ${\rm SF}_{i,{\rm M}}$ is the $i$th synthetic flux produced by model M, and $\epsilon_i$ is the excess noise term explained in Equation \ref{eq:excess}.

\begin{equation}
\label{eq:likelihood}
    \mathcal{L} = \prod_{i=1}^{N}\frac{1}{\sqrt{2\pi\epsilon_i^2}}\exp\left(-\frac{1}{2}\left(\frac{f_i - {\rm SF}_{i,{\rm M}} }{\epsilon_i}\right)^2\right)
\end{equation}

In this work, we implement BMA in two different ways to combine the posterior distribution of up to six different atmospheric models to account for possible systematic biases in the physics inputs for different models, approaching a model agnostic solution.

The first implementation, which we will refer to as averaging, consists of taking the posterior samples of each model and performing a weighted average using each model's relative probability, (computed from their evidences), as weights. The second method, which we will refer to as sampling, consists of creating a master posterior from each model's posterior samples and then re-sampling from that master posterior in order to obtain the final distributions for each parameter. This is equivalent to estimating the probability density function (PDF) of each distribution, and then performing a weighted average over them using the same methodology for the weights as previously described. These two methods use Equations \ref{eq:bma_2} and \ref{eq:bma_3} as a basis, and throughout the paper it is explicitly stated which of the two methods has been employed.

The difference between averaging and sampling is twofold: the first is the width of the credibility intervals of each parameter, with the averaging producing more precise values and the sampling generating wider intervals in general.  The second difference is important in the cases where the models themselves are in disagreement, since here the sampling will produce an estimated value drawn from the model with the highest relative probability, while the averaging produces a value that lies somewhere between the models (see Sec. \ref{sec:rad_ad} for an example).

\subsection{Parameter estimation}
\label{sec:sampling}

To estimate the best-fitting parameters we turn to nested sampling algorithms (\citealt{Skilling2004, Skilling2006}), which are designed to estimate the Bayesian evidence (marginal likelihood) of a given model, and as a by-product, the posterior distribution. These types of algorithms offer some advantages over typical Bayesian inference tools, such as MCMC algorithms (\citealt{Robert2011} and references therein).  For example, they estimate the aforementioned Bayesian evidence, they employ a well-defined stopping criterion, and provide an overall better performance in multi-modal or degenerate posterior spaces.

The evidence of a model is an essential quantity that is usually hard to calculate or estimate through either analytical or numerical methods. In problems with high dimensionality, or multimodal posteriors, there is seldom a closed analytical expression for the evidence, and it plays an important role in model selection, along with BMA (as detailed in Section \ref{sec:bma})

In order to calculate the Bayesian evidence and estimate the parameter posterior distribution with \texttt{ARIADNE} we used \texttt{MultiNest} \citep{MULTINEST} through \texttt{PyMultiNest} \citep{PYMULTINEST} for (importance) nested sampling and \texttt{dynesty} \citep{Speagle2019} for (dynamic) nested sampling \citep{Higson2019}. The main difference between regular, or 'static' nested sampling and 'dynamic' nested sampling is the ability of the latter to prioritize parameter estimation over evidence calculation. For a more in depth discussion on the differences between both methods the reader is referred to \cite{Buchner2016}.

\subsection{General algorithm}
\label{sec:algorithm}

Here we describe the main \texttt{ARIADNE} algorithm:

\begin{enumerate}
    \item If no Gaia DR2 id is provided, \texttt{ARIADNE} searches for one with the provided RA and DEC.
    \item With the Gaia DR2 id, an estimate for the T$_\text{eff}$ is queried from the Gaia DR2 catalogue. This temperature is used to select the models to fit for.
    \item A query is done to the Gaia DR2 crossmatch catalogs to see if the star is available in any of them, and if so, their respective ids are retrieved 
    \item Each catalog is queried for the corresponding ids from the previous item in order to retrieve the photometry and associated uncertainties. At the moment of retrieval, a quality control check is performed, checking each catalog's quality control flags, if any are available.
    \item If requested, a $\log{\rm g}$ estimate is calculated using MIST isochrones.
    \item The main fitting routine is executed, which consists of fitting individually and independently each selected atmospheric model.
    \item The original posterior samples are re-sampled randomly with $n$ samples, (indicated by the user), and those re-sampled posterior samples are then averaged following Sect. \ref{sec:bma}.
    \item The best fitting parameters are calculated using a kernel density estimation of the marginalized posterior samples, and the uncertainties are computed from the minimum highest density region at a 68\% significance level \citep{Hyndman1996}.
    \item Age and mass are interpolated from MIST isochrones, with the best fit parameters and photometry used as input.
\end{enumerate}

It is possible to fit single models as well, without doing BMA. In this case, however, ages and masses are not interpolated from the isochrones and only the gravitational mass is available. For more information on the gravitational mass and its comparison with the isochrone interpolated mass, the reader is referred to Section \ref{sec:mass}. \texttt{ARIADNE} can also be run in parallel with both \texttt{MultiNest} and \texttt{dynesty} through OpenMPI and native python parallelization, respectively.

We have tested \texttt{ARIADNE} on a Macbook Pro with an Intel(R) Core(TM) i7-7820HQ CPU @ 2.90GHz with 4 cores and 8 threads. Fitting individual models using \texttt{dynesty} with 500 live points and a stopping threshold of dlog z $< 0.01$, parallelized to 8 threads, takes on average five minutes; the whole BMA procedure using all six available models takes from 20 to 45 minutes depending on the star. An important note is that M dwarfs take a significant longer time to fit, meaning the BMA procedure can take up to an hour and a half with the previous settings.

\begin{figure}
    \centering
    \includegraphics[width=\columnwidth]{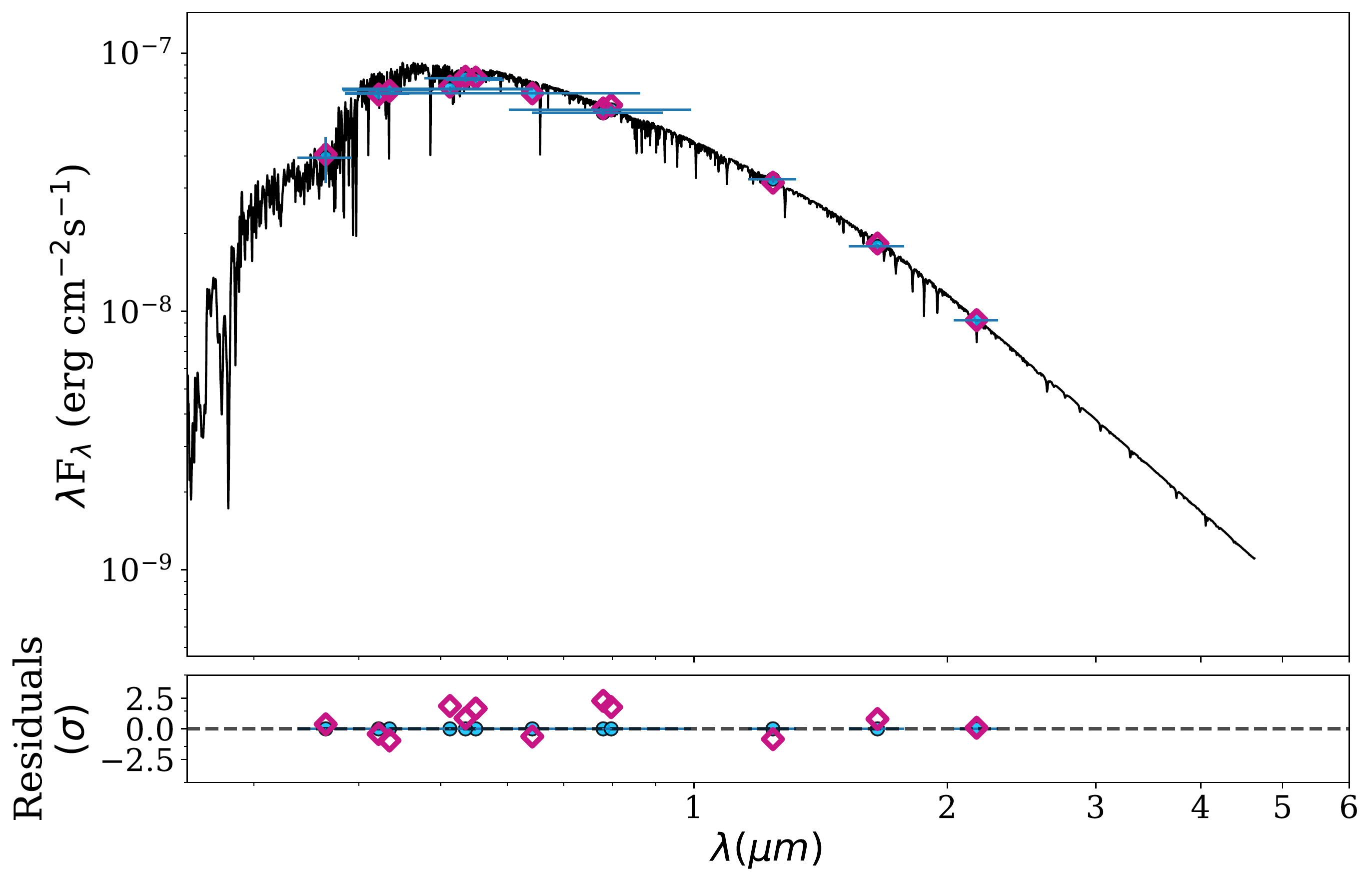}
    \caption{Top: HD 113337 BT-Settl SED plot. The blue circles are the retrieved photometry and their horizontal errorbars show the width of the filter bandpass.  The magenta diamonds are the synthetic photometry. Bottom: Residuals of the fit, normalized to the photometry errors.}
    \label{fig:SED}
\end{figure}

\begin{figure}
    \centering
    \includegraphics[width=\columnwidth]{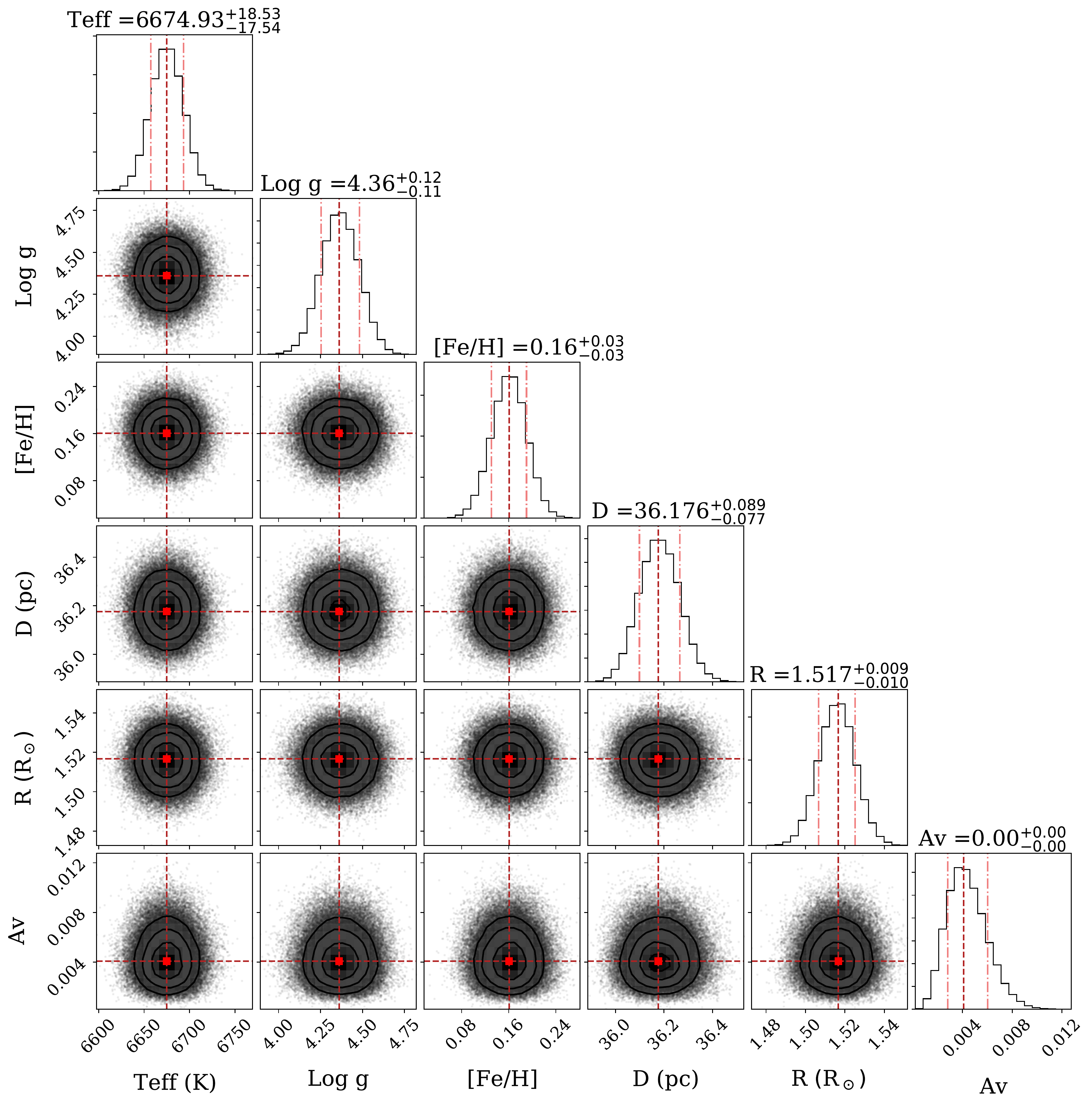}
    \caption{HD 113337 cornerplot showing the averaged distributions obtained through the averaging method.}
    \label{fig:cornerplot}
\end{figure}

\begin{figure}
    \centering
    \includegraphics[width=0.9\columnwidth]{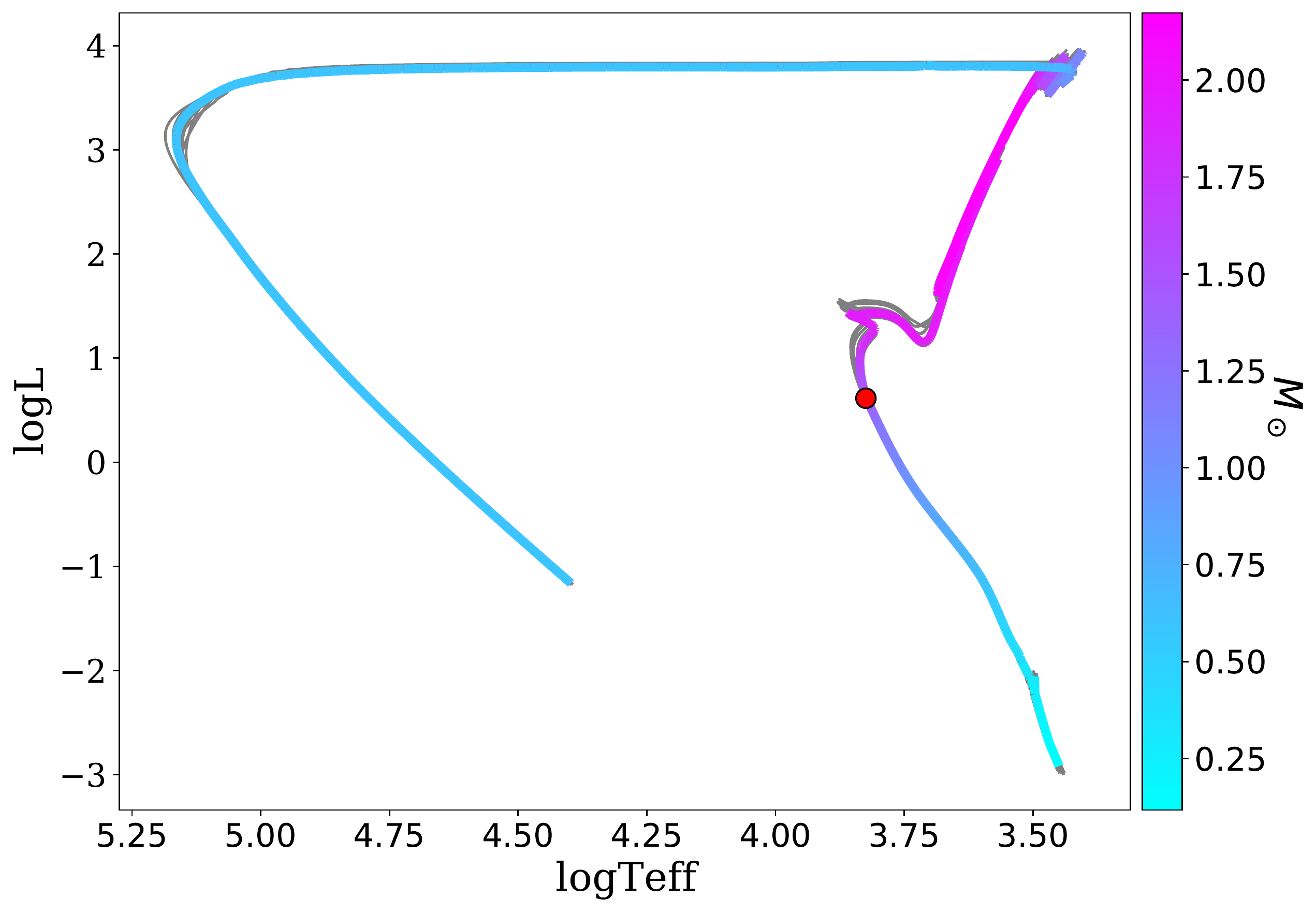}
    \caption{HD 113337 HR diagram drawn from interpolating the MIST isochrone grid to the best fitting stellar parameters. The errorbars are smaller than the symbol. The gray lines are drawn randomly from the fitted parameter samples.}
    \label{fig:HR}
\end{figure}

As standard output \texttt{ARIADNE} provides a text file containing the best fitting parameters with their associated uncertainties and their 3-$\sigma$ confidence intervals, another text file showing the photometry that was used, a third text file hosting the relative probabilities of each model, a plot of the best fitting SED model with the recovered photometry (see Sect.\ref{sec:photometry_retrieval}) and synthetic photometry, a corner plot \citep{corner} of the fitted parameters, a Hertzsprung-Russell (HR) diagram drawn from the MIST isochrones, and finally histograms showing the parameter distribution for each model. For examples of these outputs see Figures \ref{fig:SED}, \ref{fig:cornerplot}, \ref{fig:HR} and \ref{fig:rad_histograms} respectively.

\section{Benchmark stars}

\label{sec:benchmark}

To test the accuracy of \texttt{ARIADNE} we compiled a list of interferometrically observed stars, as these can provide direct measurements of the radius through the limb-darkened angular diameter and parallax distance. The selected stars cover the spectral range from A to M dwarfs, with a sub-sample of evolved stars allowing us to probe larger radii and masses, and lower $\log{\rm g}$ values. In addition to compiling interferometric radii and effective temperatures, we also searched the literature for $\log{\rm g}$ and metallicity values to compare against our outputs. In total we compiled 135 stars with radii and effective temperatures ranging from 0.15 to 16.57~R$_\odot$ and 2940 to 9377~K respectively. We list the selected stars with their respective references in Table \ref{tab:benchmark}, along with their associated magnitudes in Table \ref{tab:mags}. For each one of the benchmark stars, we ran \texttt{ARIADNE} with the default priors, except for metallicity (see Sec. \ref{sec:feh}), and using the \citet{fitzpatrick99} extinction law (testing the effects of using different extinction laws is beyond the scope of this work). We also refer the reader to other published works that have made use of \texttt{ARIADNE} to calculate stellar parameters for stars up to 3 kpc away (e.g. see \citealt{Diaz, James, Jack, Rosie, Smith}).

\begin{table*}
\caption{Benchmark stars used to test \texttt{ARIADNE}. A full version is available as supplementary material in machine-readable format. The first rows are provided for guidance}              
\label{tab:benchmark}      
\centering   
\begin{tabular}{c c c c c c c c}          
\hline\hline                        
Star & T$_\text{eff}$ & $\log{\rm g}$ & [Fe/H] & $\theta_\text{LD}$ & Radius    & $\log~R_{\rm HK}$ & Refs \\
Name &    K           &  dex          & dex    & mas                & R$_\odot$ & dex               &      \\
\hline 
HD166    & 5327 $\pm$ 39  & 4.58 $\pm$ 0.11 & 0.08 $\pm$ 0.11  & 0.624 $\pm$ 0.009  & 0.917 $\pm$ 0.009 & -4.36 & 27,19 \\
HD48737  & 6480 $\pm$ 39  & 3.61 $\pm$ 0.10 & -0.09            & 1.401 $\pm$ 0.009  & 2.710 $\pm$ 0.021 & -4.03 & 20,21 \\
\vdots & \vdots & \vdots & \vdots & \vdots & \vdots & \vdots & \vdots \\
\hline \\
\multicolumn{8}{l}{\textbf{Notes.} $\log R_{\rm HK}$. values from \citet{Boro2018}} \\
\multicolumn{8}{l}{Assumed 10\% uncertainties on values where there are none reported.} \\
\multicolumn{8}{l}{\textbf{References.} (1) \citet{Rains2020}; (2) \citet{Delgado2017}; (3) \citet{Jenkins2011}; (4) \citet{Blackwell1998}} \\
\multicolumn{8}{l}{(5) \citet{Erspamer2003}; (6) \citet{Royer2007}; (7) \citet{Martinez2010}; (8) \citet{Ramirez2013}; (9) \citet{Massarotti2008}} \\
\multicolumn{8}{l}{(10) \citet{Allende2008}; (11) \citet{Alves2015}; (12) \citet{Liu2007}; (13) \citet{Hekker2007}; (14) \citet{Torres2006}} \\
\multicolumn{8}{l}{(15) \citet{Boyajian2012b}; (16) \citet{Fischer2005}; (17) \citet{Casagrande2011}; (18) \citet{Mallik2003}; (19) \citet{Soubiran2016}} \\
\multicolumn{8}{l}{(20) \citet{Boyajian2012}; (21) \citet{Gray2003}; (22) \citet{Anderson2011}; (23) \citet{vonBraun2014}; (24) \citet{Rojas-Ayala2012}} \\
\multicolumn{8}{l}{(25) \citet{Torres2008}; (26) \citet{Boyajian2015}; (27) \citet{Boyajian2013}; (28) \citet{Borgniet2019}} \\
\multicolumn{8}{l}{(29) \citet{Ligi2019}; (30) \citet{Brewer2016}; (31) \citet{Mortier2013}; (32) \citet{White2018}} \\
\multicolumn{8}{l}{(33) \citet{Tanner2015}; (34) \citet{Lachaume2019}; (35) \citet{Houdebine2019}; (36) \citet{Stokholm2019}} \\
\multicolumn{8}{l}{(37) \citet{Johnson2014}; (38) \citet{Ligi2016}; (39) \citet{Berger2006}; (40) \citet{Huber2016}} \\
\multicolumn{8}{l}{(41) \citet{Ammons2006}; (42) \citet{Bourges2017}; (43) \citet{Bigot2011}; (44) \citet{vonBraun2011}} \\
\multicolumn{8}{l}{(45) \citet{Santos2004}; (46) \citet{Baines2008}; (47) \citet{Mozurkewich2003}; (48) \citet{Huber2012}} \\
\multicolumn{8}{l}{(49) \citet{Crepp2012}; (50) \citet{Nordgren1999}; (51) \citet{vanBelle2009}; (52) \citet{Baines2012}} \\
\end{tabular}
\end{table*}

\begin{table*}
\caption{Magnitudes of the benchmark stars used to test \texttt{ARIADNE}. A full version is available as supplementary material in machine-readable format. The first rows and columns are provided for guidance}    
\begin{tabular}{cccccccc}
\hline\hline
Star Name & 2MASS\_H & 2MASS\_H\_e & 2MASS\_J & 2MASS\_J\_e & 2MASS\_Ks & 2MASS\_Ks\_e & \dots \\
\hline
GJ1139   &   6.707 &     0.020 &   7.315 &     0.021 &    6.594 &      0.021 & \dots\\
GJ1189   &   6.841 &     0.020 &   7.385 &     0.027 &    6.742 &      0.029 & \dots\\
GJ15A    &   4.476 &     0.200 &   5.252 &     0.264 &        - &          - & \dots\\
GJ166A   &   2.594 &     0.198 &   3.013 &     0.238 &    2.498 &      0.236 & \dots\\
\vdots   & \vdots  &    \vdots &  \vdots &    \vdots &   \vdots &     \vdots & \dots\\
\hline
\end{tabular}
\label{tab:mags}
\end{table*}

\section{Results}
\label{sec:results}
In the following section we discuss the results obtained when running \texttt{ARIADNE} on the benchmark stars, investigating the precision of the code for the main atmospheric parameters: T$_{\rm eff}$, $\log{\rm g}$, [Fe/H], radius, and mass. We have compiled all of \texttt{ARIADNE}'s outputs in Table \ref{tab:output}.

\begin{table*}
\caption{\texttt{ARIADNE} outputs for the benchmark stars with their 1-$\sigma$ errorbars. A full version is available as supplementary material in machine-readable format. The first rows are provided for guidance.}              
\label{tab:output}      
\centering   
\begin{tabular}{c c c c c c c c c}          
\hline\hline                        
Star &  T$_\text{eff}$ & $\log{\rm g}$ & [Fe/H] & Radius  & Mass$_\text{grav}$ & Mass$_\text{iso}$ & $\theta_\text{LD}$ & A$_{\rm V}$\\
Name & K & dex & dex & R$_\odot$ & M$_\odot$ & M$_\odot$ & mas & mag\\
\hline 
HD 209458 & $5994^{+13}_{-7}$ & $4.38^{+0.03}_{-0.07}$ & $0.19^{+0.02}_{-0.01}$ & $1.213^{+0.009}_{-0.007}$ & $1.28^{+0.09}_{-0.22}$ & $1.15^{+0.01}_{-0.01}$ & $0.2328^{+0.0018}_{-0.0020}$ & $0.005^{+0.001}_{-0.003}$ \vspace{2pt} \\
HD 189733 & $5049^{+19}_{-16}$ & $4.51^{+0.09}_{-0.10}$ & $-0.03^{+0.03}_{-0.04}$ & $0.765^{+0.007}_{-0.006}$ & $0.66^{+0.10}_{-0.21}$ & $0.81^{+0.01}_{-0.01}$ & $0.3593^{+0.0025}_{-0.0038}$ & $0.002^{+0.000}_{-0.005}$ \vspace{2pt} \\
HD 113337 & $6675^{+18}_{-19}$ & $4.36^{+0.11}_{-0.12}$ & $0.16^{+0.03}_{-0.03}$ & $1.517^{+0.010}_{-0.009}$ & $1.79^{+0.29}_{-0.75}$ & $1.40^{+0.01}_{-0.02}$ & $0.3898^{0.0026}_{-0.0025}$ & $0.004^{+0.001}_{-0.002}$ \vspace{2pt} \\
\vdots & \vdots & \vdots & \vdots & \vdots & \vdots & \vdots & \vdots \\
\hline \\
\end{tabular}
\end{table*}

\subsection{Radius accuracy}
\label{sec:rad_ad}

\begin{figure}
\label{fig:rad_ad}
\centering

\subfloat[\label{fig:rad_comp} The \texttt{ARIADNE} $R_\star$ from the averaging method compared to computed interferometric radii from angular diameters and BJ18 distances, color coded with $\log R_{HK}$. The red dots are stars without a $\log R_{HK}$ measurement. The dashed orange line represents the 1:1 relation. The lower panel shows the fractional residuals which indicate a rising trend for the smallest stars.]{\includegraphics[width=0.97\columnwidth]{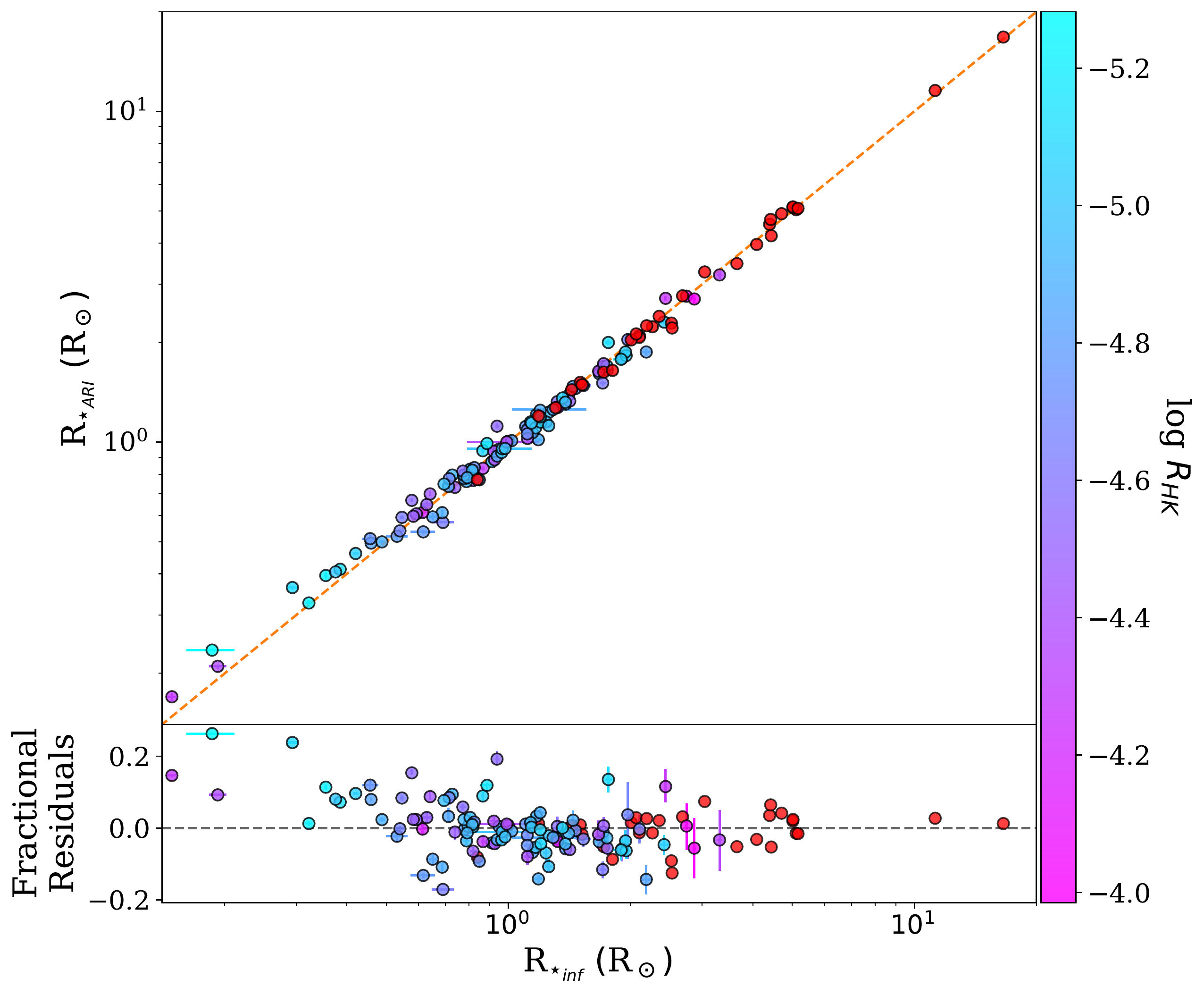}%
}

\subfloat[\label{fig:ad_comp} Computed angular diameters from the \texttt{ARIADNE} $R_\star$ computed with the averaging method shown in (a) and the distance outputs compared to interferometric observations of angular diameters from the benchmark stars. The dashed orange line represents the 1:1 relation. The lower panel shows the fractional residuals, which, as opposed to the radius plot, does not show significant behavior with varying angular diameter.]{\includegraphics[width=0.97\columnwidth]{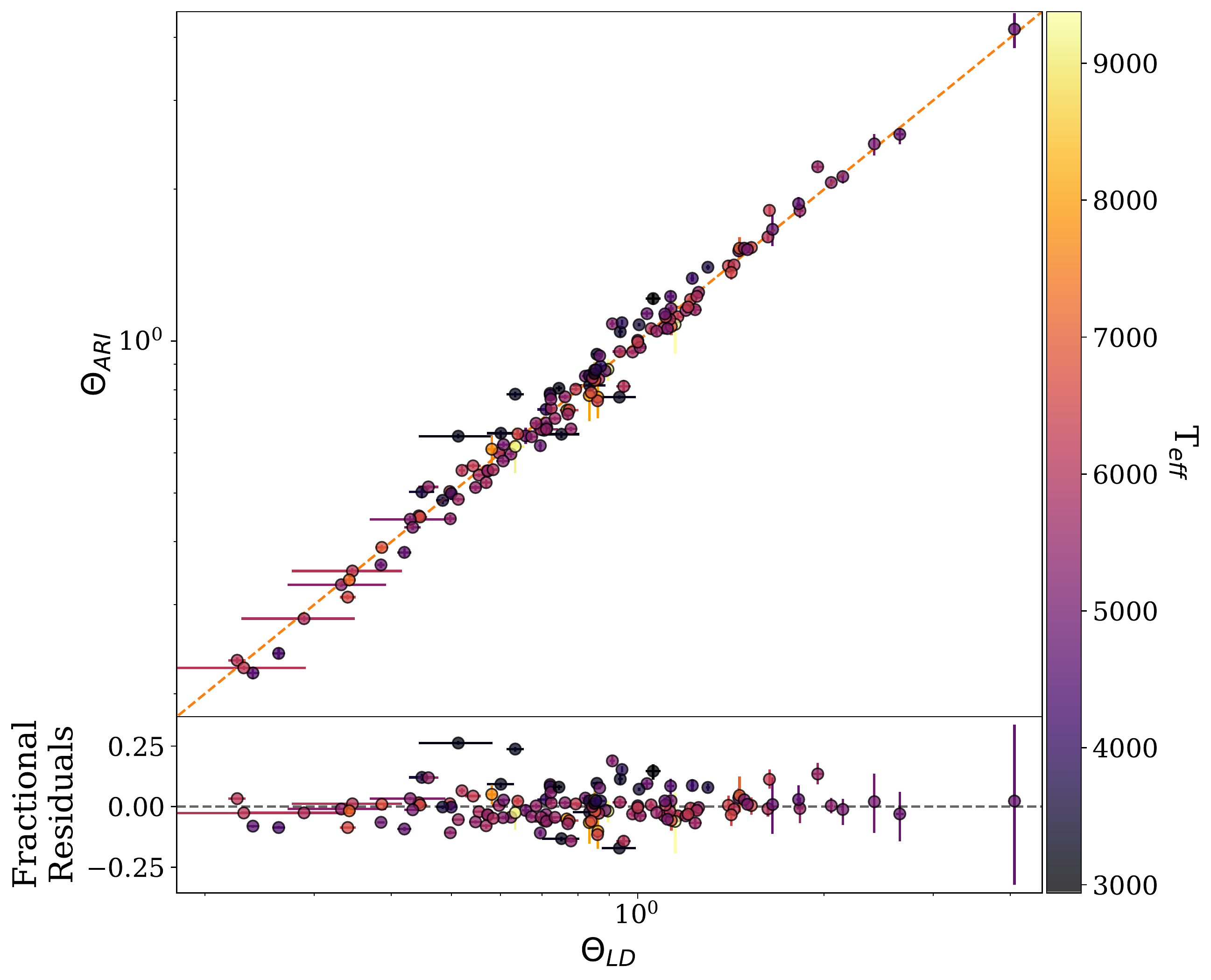}%
}

\caption{Comparison between literature radii and angular diameters vs our outputs.}

\end{figure}

Due to the nature of the model, the radius correlates directly with the distance through the normalization factor defined in Section \ref{sec:convolution}. The distance is tightly constrained by the BJ18 derived distances, thus giving highly constrained values for $R_\star$, even when using uninformative priors. To highlight this effect, we ran \texttt{ARIADNE} on the planet hosting star HD 113337 using both the BJ18 distance and Hipparcos parallax as priors for the distance to the system.  We found radii of $1.52\pm0.01~R_{\odot}$ and $1.58^{+0.05}_{-0.04}$~R$_\odot$ for the BJ18 distance and Hipparcos' distance, respectively, revealing an offset at a level higher than 1-$\sigma$ from the BJ18 value, as well as uncertainties three to five times larger. Since most of the stars listed in Table \ref{tab:benchmark} had their radii calculated using the parallax from Hipparcos \citep{VanLeeuwen2007}, which are significantly less well constrained than Gaia, and in most cases have an offset, we re-calculated the radii of the stars using the BJ18 distance and compared those values to \texttt{ARIADNE}'s output.  In Figure \ref{fig:rad_comp}, we highlight that the interferometrically observed radii and our measured values follow a 1:1 relation, with fractional RMS values of 0.001$\pm$0.070 and 0.000$\pm$0.069, and mean precisions of $2.1\%$ and $4.3$\% for the averaging and sampling methods, respectively.
While the expected scatter derived from interferometrically observed stars ranges from 1.5\% for A, F and G stars to 5\% for K and M stars, we chose the more conservative value of 5\% for comparison \citep{Stassun2017}, and find that the intrinsic scatter computed from \texttt{ARIADNE} lies well within the interferometric scatter. We find an increasing trend for the smallest stars, those with $R_\star \lesssim 0.4-0.5$R$_\odot$, (the M dwarfs), and therefore it appears more work is necessary to properly model these objects, both in the community and within ARIADNE itself.  However, we also note that the slope does not cross the blue interferometric scatter region, showing that the effect is small in reality.

For comparisons sake, we also compute angular diameters for the benchmark stars from \texttt{ARIADNE} derived $R_\star$ and distance outputs (recall the angular diameter definition: $\Theta\equiv 2R_\star/d$) and compared them to the interferometrically observed angular diameters.  Our results show that the angular diameters derived from the SED fit correlates well with the observational counterpart, as shown in Figure \ref{fig:ad_comp}.

\begin{figure*}

\subfloat[\label{fig:sub-first} HD 222603. Best case scenario where all models agree and the probabilities are approximately equally distributed. The dot-dashed line represents the averaged distributions while the dashed line shows the distribution drawn from the weighted sampling. This applies for all panels in this Figure.]{\includegraphics[width=1\columnwidth]{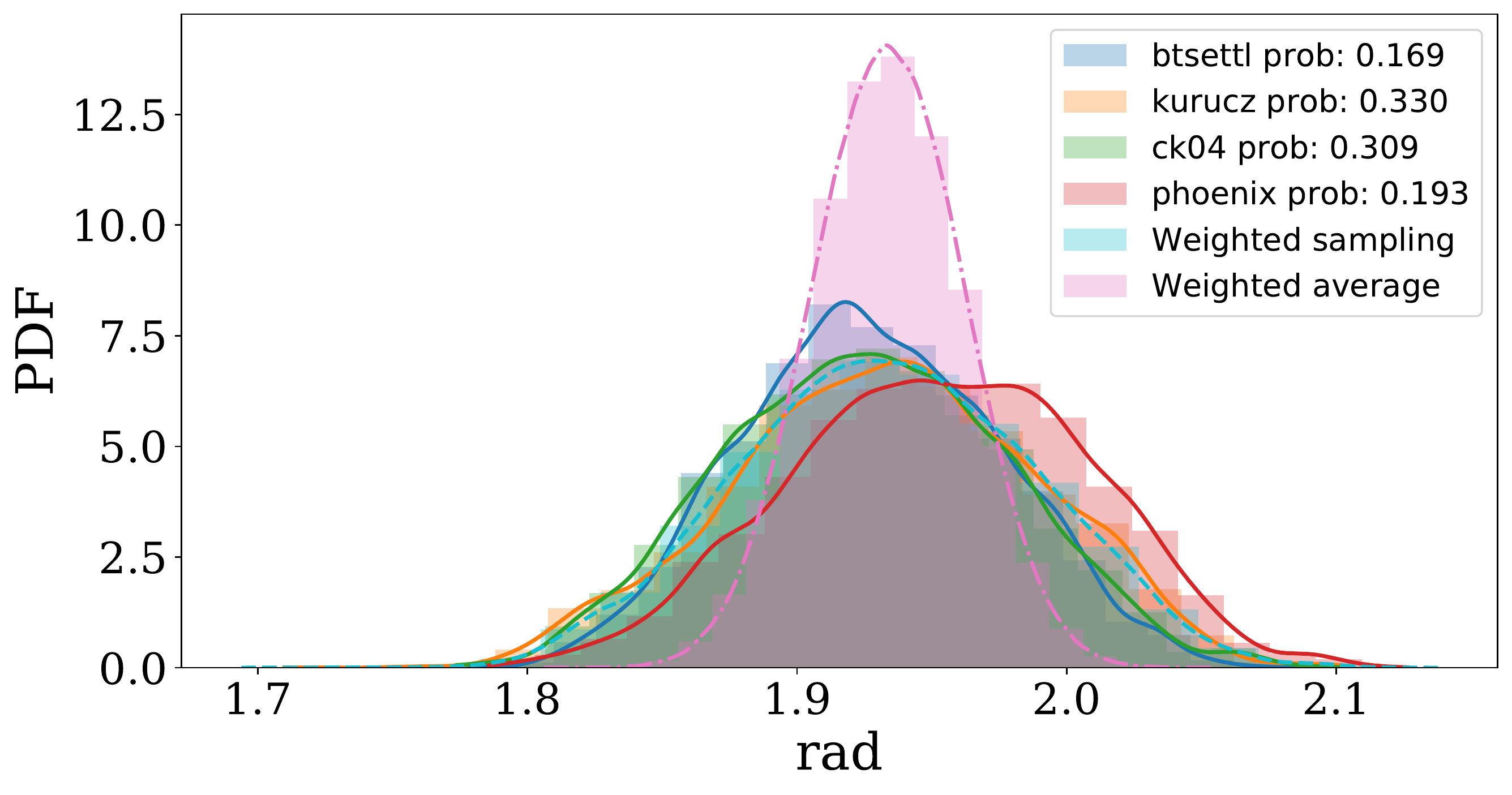}%
}\hspace{8pt}%
\subfloat[\label{fig:sub-second} HD 143761. There are two groups of posteriors, one with the PHOENIX v2 and Castelli \& Kurucz, and another with BT-Settl and Kurucz models.]{\includegraphics[width=1\columnwidth]{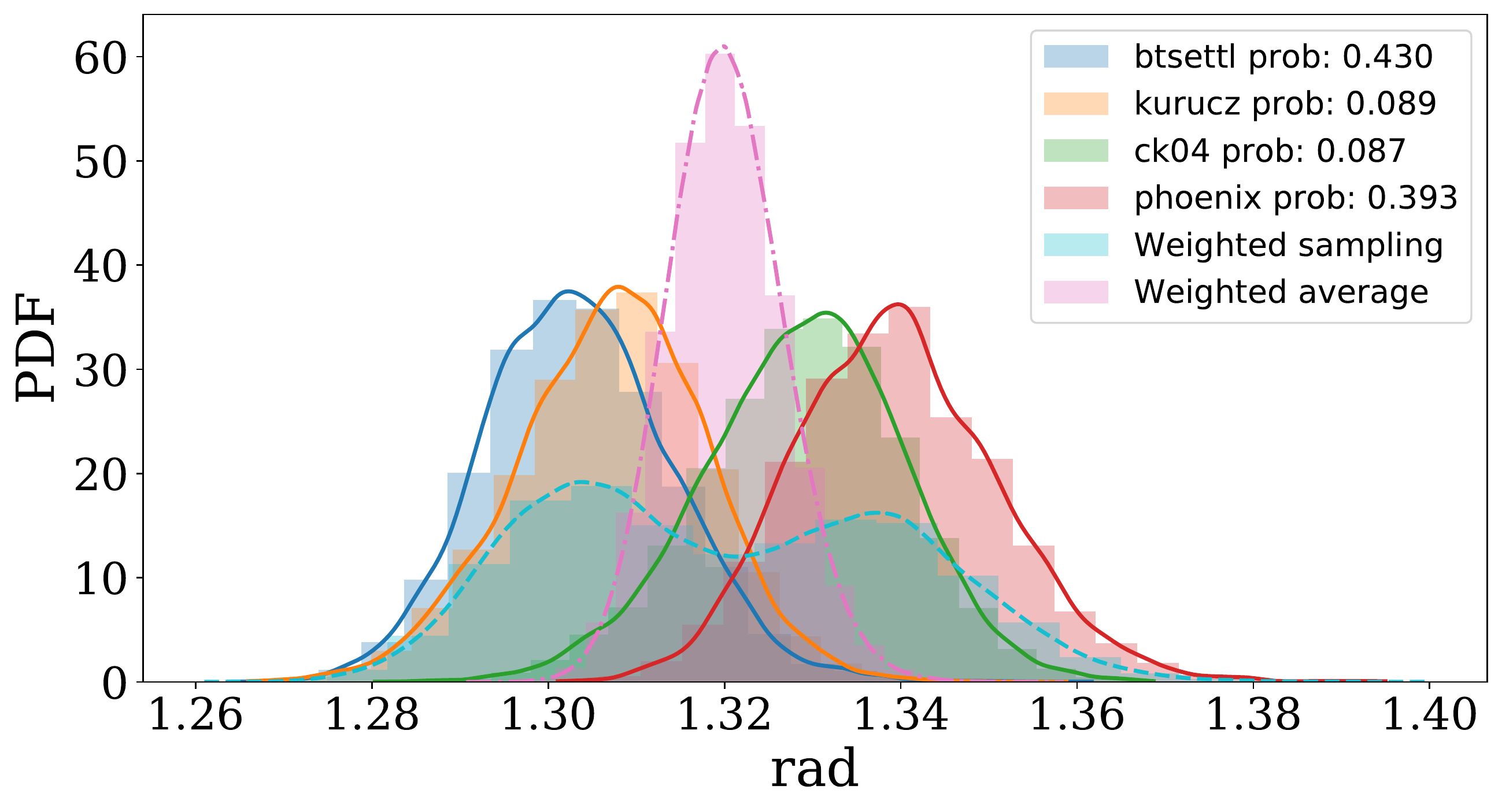}%
}
  
\subfloat[\label{fig:sub-third} GJ 33. The posterior samples are, in general, separated and each model has a similar probability.]{\includegraphics[width=1.05\columnwidth]{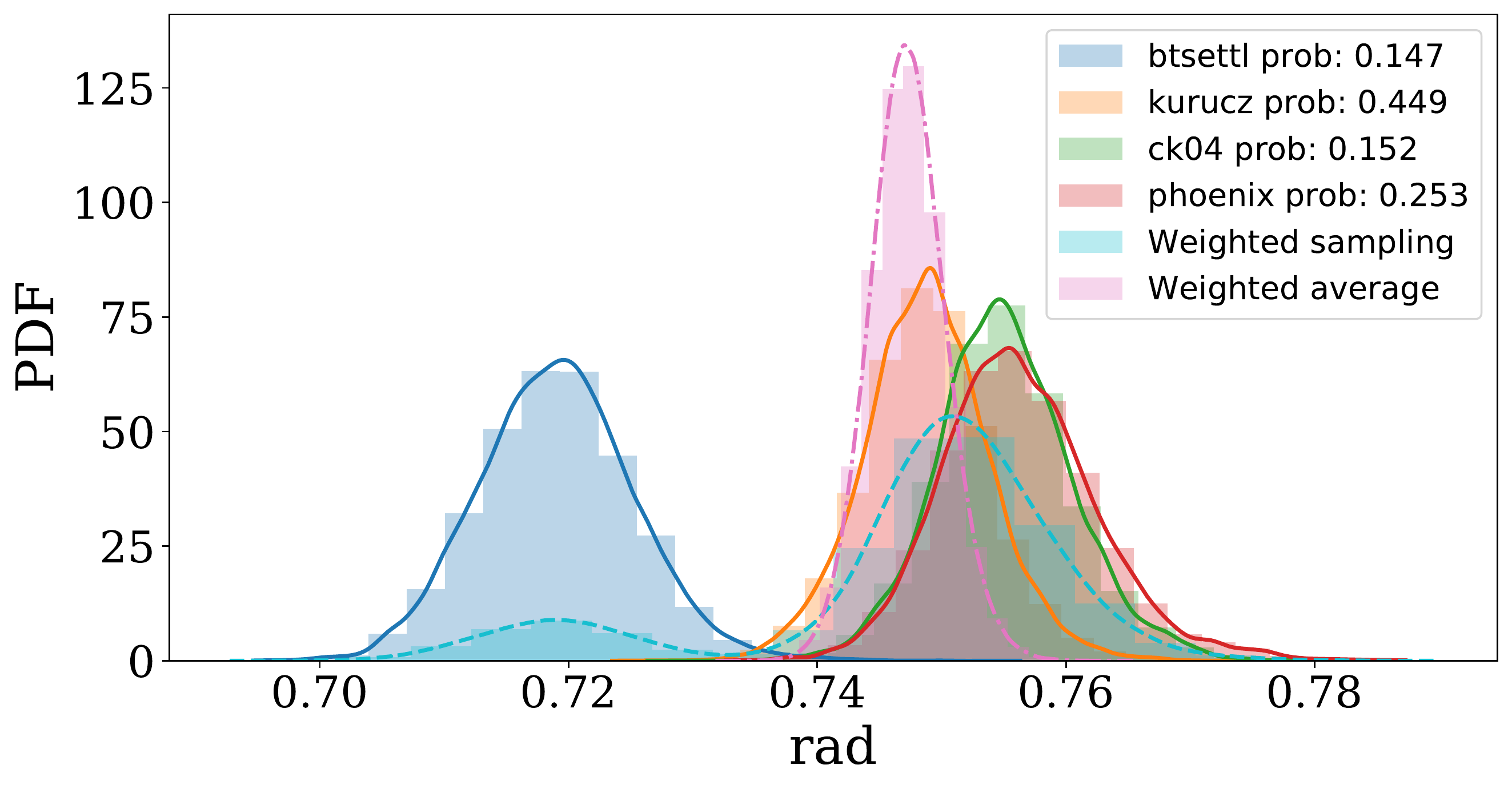}%
}\hspace{8pt}%
\subfloat[\label{fig:sub-fourth} HD 170493. There is a single dominant model. Note that in this case the model averaging reduces to model selection and thus the averaged histogram is nearly identical to the dominant model.]{\includegraphics[width=1\columnwidth]{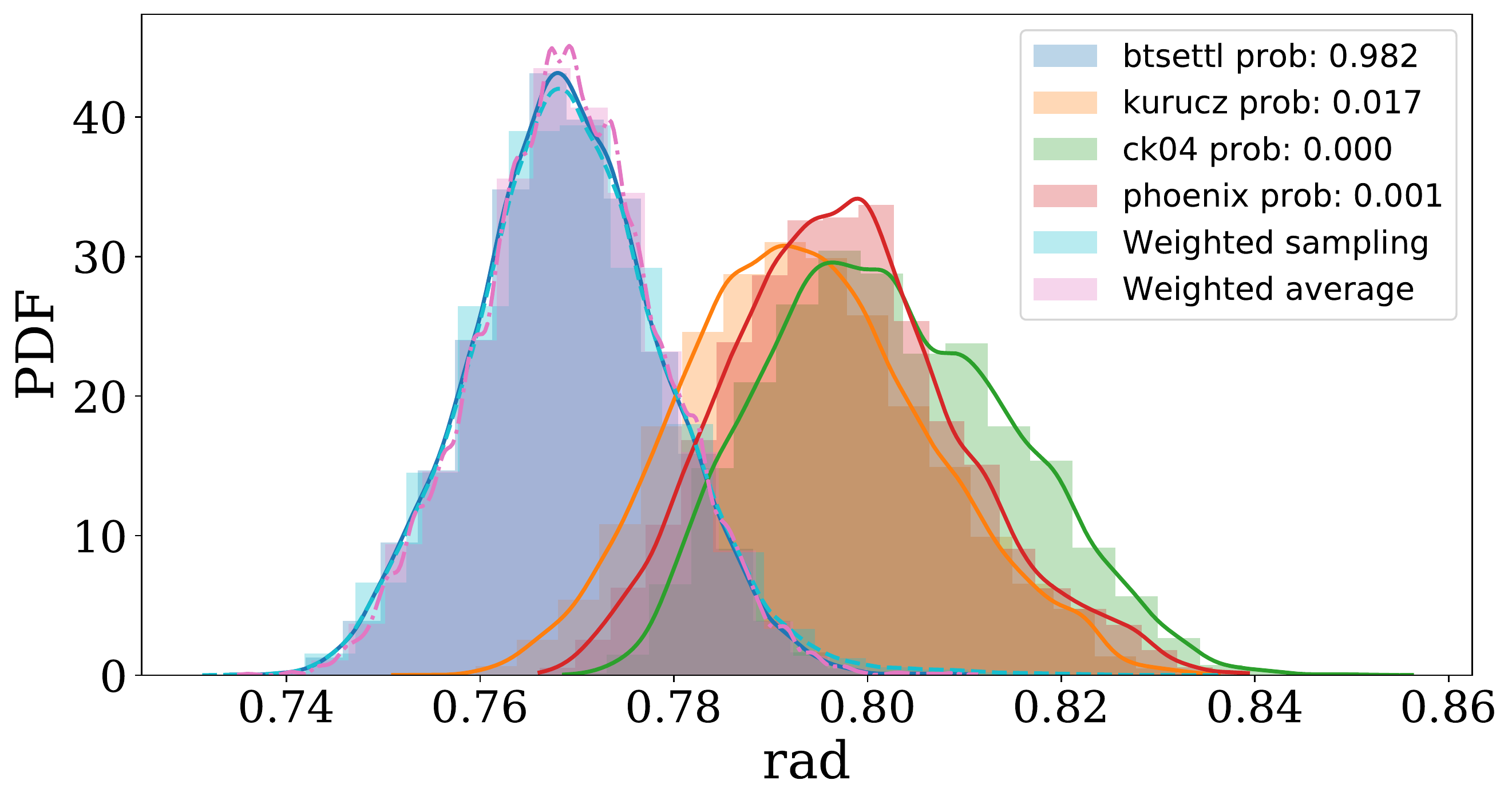}%
}
\caption{Radius posterior samples for different stars with each model's probability and the averaged posterior highlighting various possible scenarios when modeling stars with \texttt{ARIADNE}.}
\label{fig:rad_histograms}

\end{figure*}

\begin{figure}
\centering
 \includegraphics[width=1\columnwidth]{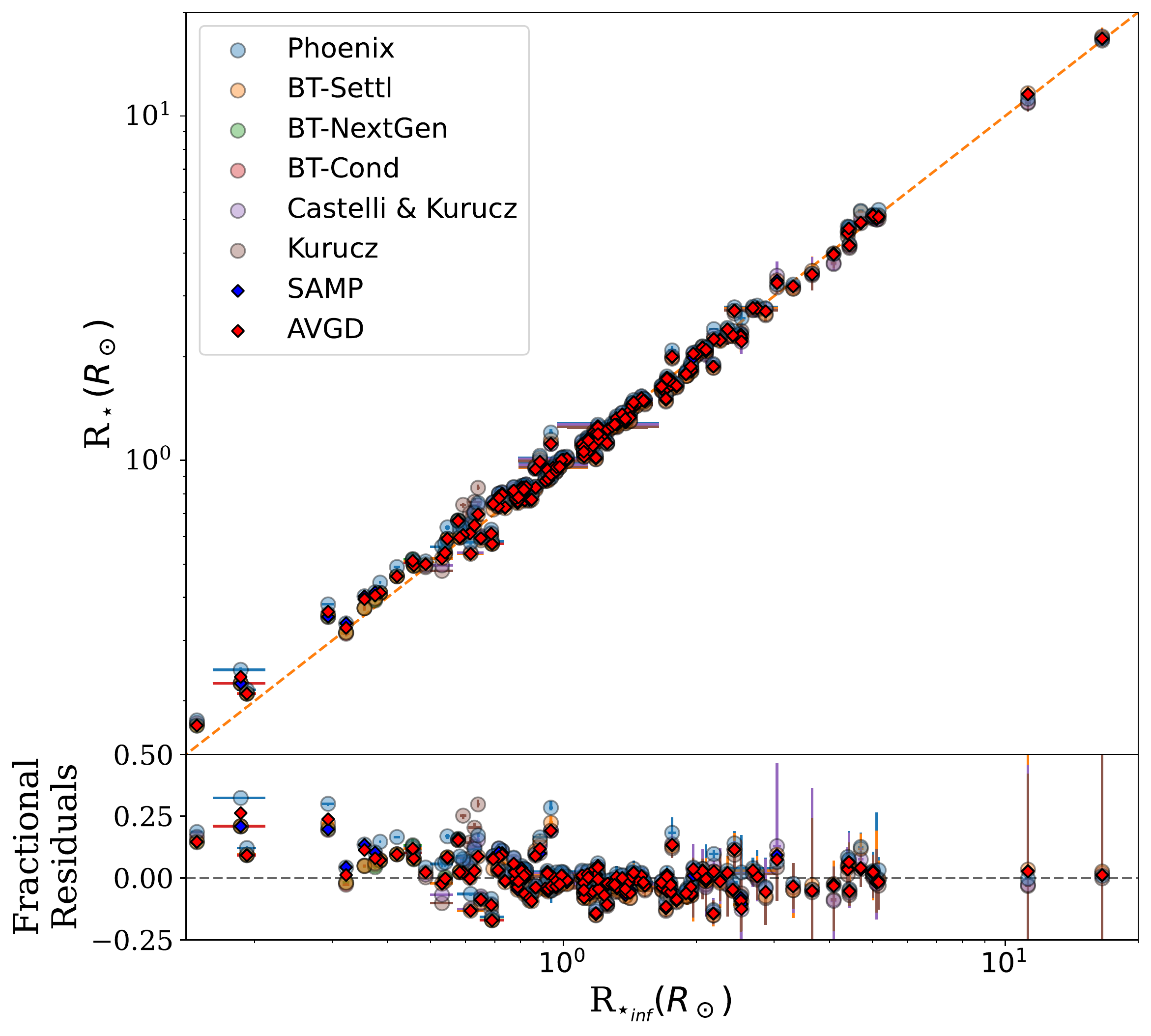}
 \caption{Radii calculated with \texttt{ARIADNE} compared against the interferometric values. Individual model estimates are shown, along with the averaging values in red diamonds and the sampling values in blue diamonds. It can be seen that in most cases the averaged values are more accurate than the individual models. The lower panel shows the fractional residuals where it can be seen that the most accurate values lie in the 0.75-1.50 R$_\odot$ range, and the rising trend for the smallest stars ($R_\star < \sim$0.4-0.5~R$_\odot$) is also apparent.}
 \label{fig:all_rad}
\end{figure}

To show the effect BMA and proper weighting has on the precision of the stellar parameters, in particular for radii, we provide histograms showing the distribution of each model, alongside the probability associated with them, and the distribution of the averaged posterior samples, all part of the \texttt{ARIADNE} standard output.  An example of these histograms are shown in Figure \ref{fig:rad_histograms}. In addition, Figure \ref{fig:all_rad} shows the effect averaging has on the final radius value in the sample. We note that there are four main behaviors for the radii posteriors. The best case scenario is that all of the model posteriors agree within 1-$\sigma$, case highlighted in Figure \ref{fig:sub-first}. Figure \ref{fig:sub-second} shows the case where there are two distinct groups (separated by more than 2-$\sigma$) where the posteriors are clustered. The third case, shown in Figure \ref{fig:sub-third}, is where the models are separated across a wide area of the parameter space, with no dominant single model. The second and third cases are where BMA has the largest impact, (although in the first case, where all models agree within 1-$\sigma$, BMA further constrains the final posterior distribution giving a more precise value), unifying the conflicting models by weighting them by their according probabilities. Figure \ref{fig:sub-fourth} is the fourth scenario where a single model dominates over the rest, and in this case BMA is performing model selection.

\subsection{Teff accuracy}
\label{sec:teff}

\begin{figure}
\centering
 \includegraphics[width=1\columnwidth]{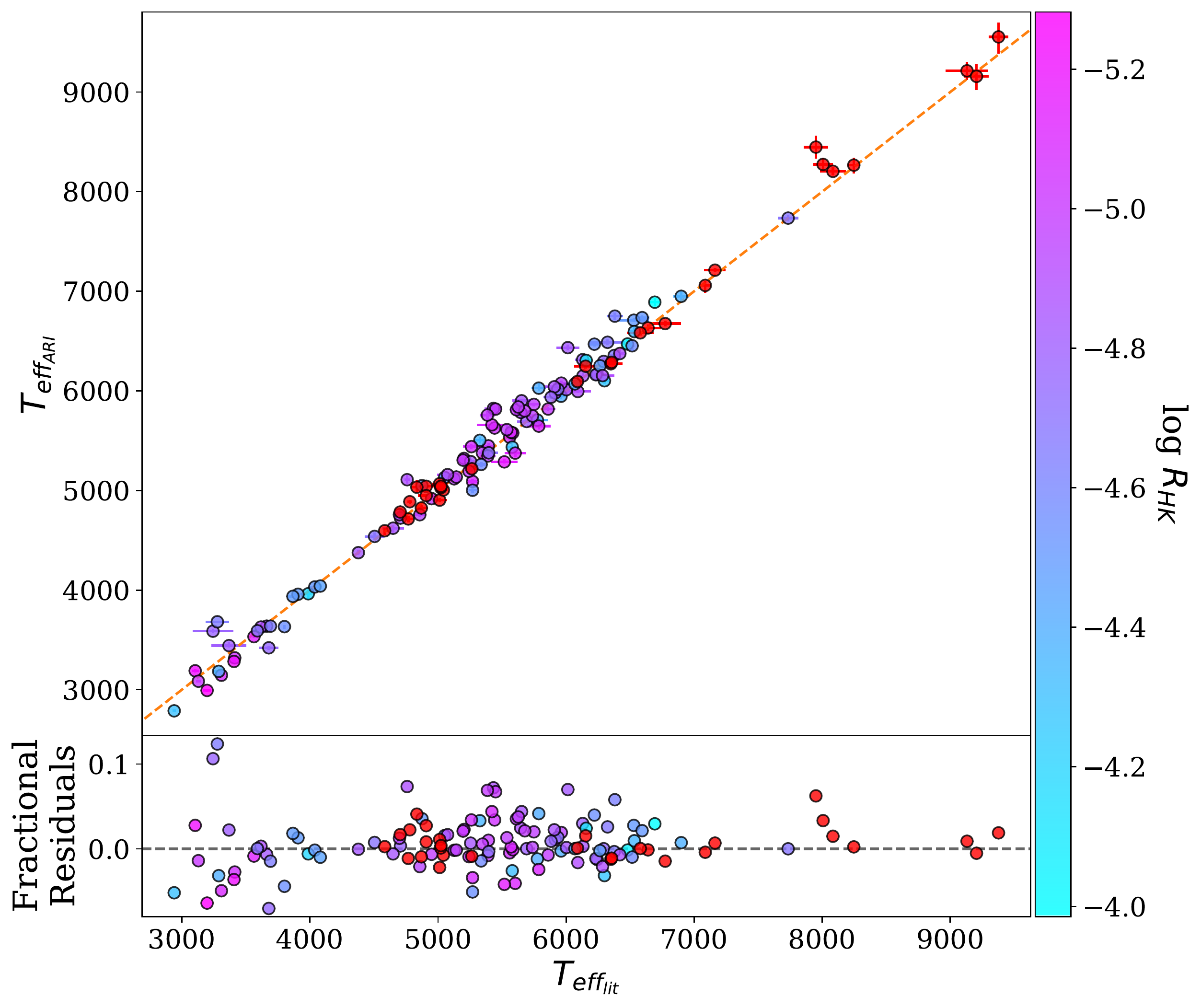}
 \caption{\texttt{ARIADNE} T$_\text{eff}$ from the averaging method output compared against literature values for the benchmark stars. The dashed blue line represents the 1:1 relation. The lower panel shows the residuals.}
 \label{fig:teff_comp}
\end{figure}

The effective temperature of a star is a key parameter, as it is one of the main determinants in the overall shape of the SED and thus can be precisely determined from SED fits alone, mostly with uncertainties smaller than $\pm$100 K.  Effective temperature also helps to further constrain the radius. We found that \texttt{ARIADNE} reproduces the T$_{\rm eff}$ literature values accurately. We show this comparison in Figure \ref{fig:teff_comp}. We also tested varying the priors for the temperature between 2-10 times the error reported and adding $\pm$100~K to the value reported by the literature to test the effect that wider priors have on the precision of the fit.  In these cases, we found that there is a slight decrease in precision of around $\pm$10~K when using an extremely wide prior, such as the RAVE prior (see Sect. \ref{sec:priors}), otherwise \texttt{ARIADNE} is insensitive to T$_{\rm eff}$ prior choice. We also note that using a wider prior for the temperature decreases the radius precision by around a factor of two.

\subsection{log g accuracy}
\label{sec:logg}

\begin{figure}
\centering
 \includegraphics[width=1\columnwidth]{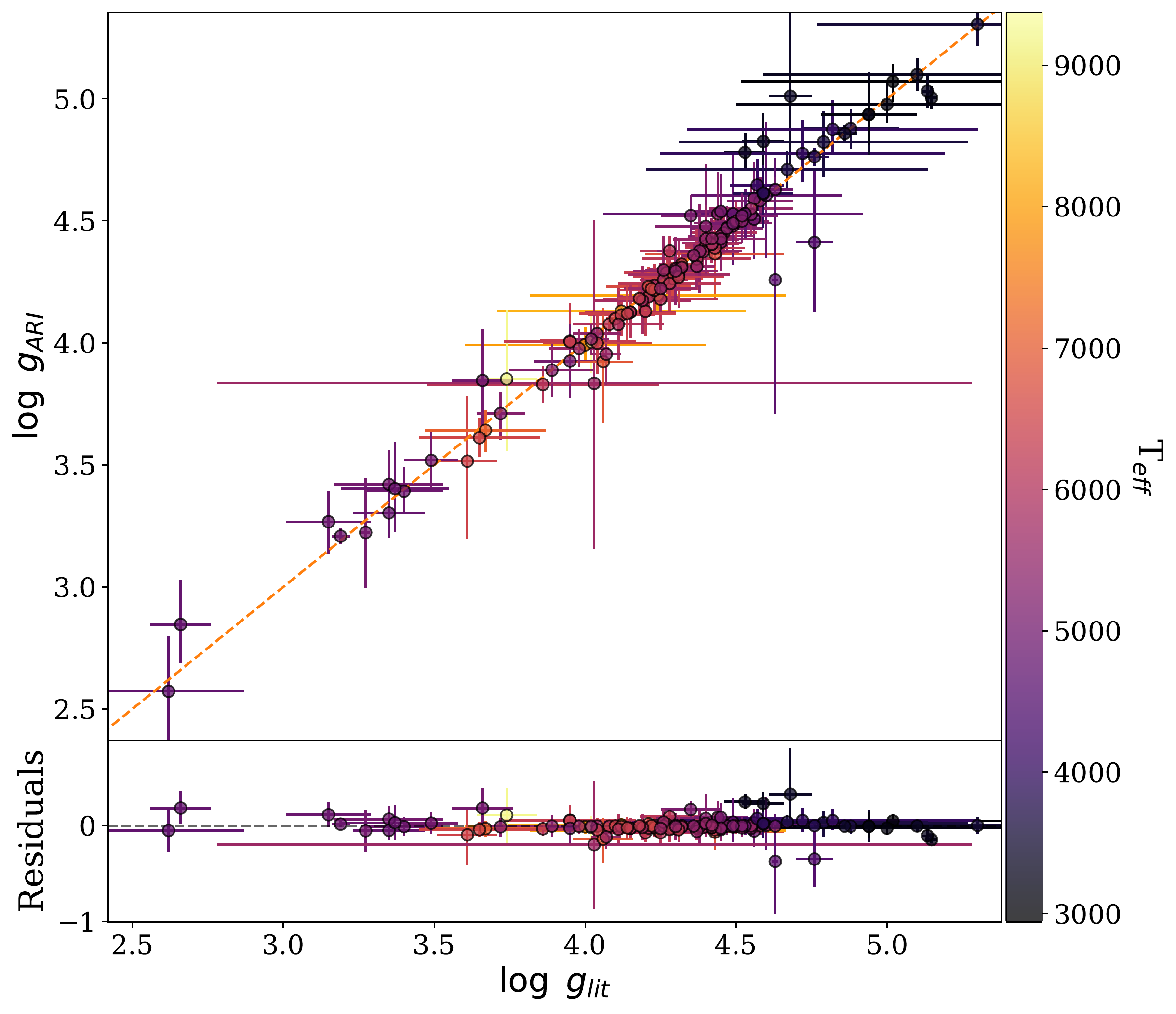}
 \caption{\texttt{ARIADNE} $\log{\rm g}$ from the averaging method output compared against literature values for the benchmark stars. The dashed orange line represents the 1:1 relation. The lower panel shows the residuals.}
 \label{fig:logg_comp}
\end{figure}

The surface gravity is hard to capture from SEDs alone, and using a global prior drawn from RAVE tends to consistently underestimate the $\log{\rm g}$.  Therefore, we recommend drawing a prior from an independent method, such as spectral analysis. There is an important reservation regarding using evolutionary codes, such as MIST, for the estimation of the $\log{\rm g}$, since they could possibly use, or incorporate, archival photometry, which could have the knock-on effect of arriving in a position where data has been used twice when running the analysis with \texttt{ARIADNE}. For the analysis done on the benchmark stars, the $\log{\rm g}$ prior was taken from the literature values, and whenever the $\log{\rm g}$ was missing, a Uniform prior between 3.5 and 6.0 was used instead. Figure \ref{fig:logg_comp} shows the comparison with the literature values excluding those stars where no literature $\log{\rm g}$ could be found.

\subsection{Metallicity}
\label{sec:feh}

\begin{figure}
\centering
 \includegraphics[width=1\columnwidth]{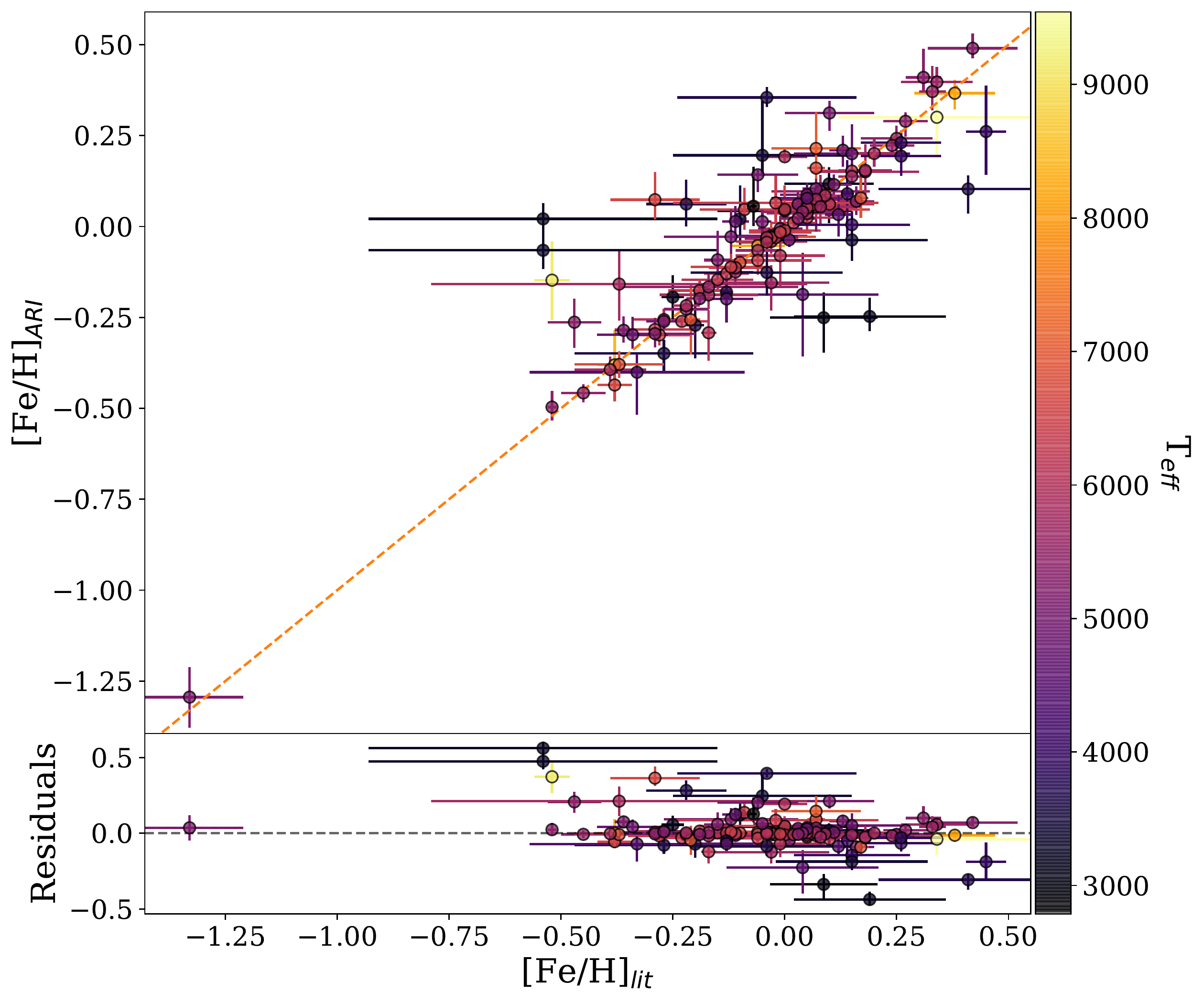}
 \caption{\texttt{ARIADNE}'s [Fe/H] from the averaging method output compared against literature values for the benchmark stars. Each star is color coded by it's effective temperature, taken from the literature. The dashed orange line represents the 1:1 relation. The lower panel shows the residuals.}
 \label{fig:metal_comp}
\end{figure}

The metallicity cannot be consistently constrained through our SED fitting procedure when using an unconstrained prior such as the one drawn from RAVE (see Sect. \ref{sec:priors}). This is mainly because the metallicity has a negligible effect on the overall shape of the SED and instead affects narrow absorption lines, something that is not effectively captured by broadband photometry. We also tested including narrow-band photometric measurements in the fitting procedure, in particular the Str\"omgren filters that were designed to measure metallicities of stars, however these also proved ineffective in increasing the accuracy of our metallicity measurements. Including spectroscopically derived priors drastically improves the accuracy for the metallicity, so we drew priors from the PASTEL catalog \citep{Soubiran2016} when available. Figure \ref{fig:metal_comp} shows our metallicity results using this prior.

\subsection{Mass calculation}
\label{sec:mass}

We offer two methods of calculating stellar masses with \texttt{ARIADNE}. The first is the so-called gravitational mass, $M$ in units of the Solar mass, which is calculated directly from the star's $\log{\rm g}$ and $R$ in units of Solar radius, using equation \ref{eq:grav_mass}

\begin{equation}
\label{eq:grav_mass}
    g = g_\odot\frac{M}{R^2}
\end{equation}
which after taking the logarithm becomes

\begin{equation}
\label{eq:grav_mass_log}
    \log{\rm g}=\log~M-2\log~R+4.437
\end{equation}

The second method involves interpolating MIST isochrones to the previously fitted stellar parameters and broadband photometry, however this option is only available when doing the BMA procedure and not for individual model fitting. 

\section{what makes ARIADNE unique?}
\label{sec:vosa}

As mentioned in the introduction, a number of algorithms and methodologies have been developed to measure stellar bulk properties, however only one employs Bayesian statistics \citep{Gillen2017}, and they only apply the analysis to an individual model set.  VOSA on the other hand, is a public web-tool designed to build SEDs and obtain different properties from Virtual Observatory catalogs, by modeling SEDs with synthetic photometry using a $\chi^2$ minimization grid \citep{Bayo2008}. Given that VOSA has been updated to include a "Bayes analysis", which consists of doing the $\chi^2$ minimization routine and then assigning relative probabilities to each model, finally normalizing them, we may ask the question, why is \texttt{ARIADNE} necessary?  Does VOSA not accomplish the same outcomes as \texttt{ARIADNE}?

The VOSA procedure produces relative probabilities of each point in the selected model grids, but does not include priors for the analysis, which prove to be even more fundamental in SED analyses since the shape of the SED is not enough to constrain some of the parameters (e.g. log~g and [Fe/H]). Another important difference is that VOSA evaluates the $\chi^2$ only on the model grid points, while \texttt{ARIADNE} interpolates between those points to take full advantage of the exploratory nature of the nested sampling algorithm. Therefore, and most crucially, VOSA can not make use of the power of BMA when determining the stellar parameters, and indeed BMA is not an option within the VOSA package, nor any other current package that performs these types of analyses.

\section{Model offsets and discussion}
\label{sec:offset}

\begin{figure*}
    \centering
    \includegraphics[scale=.6]{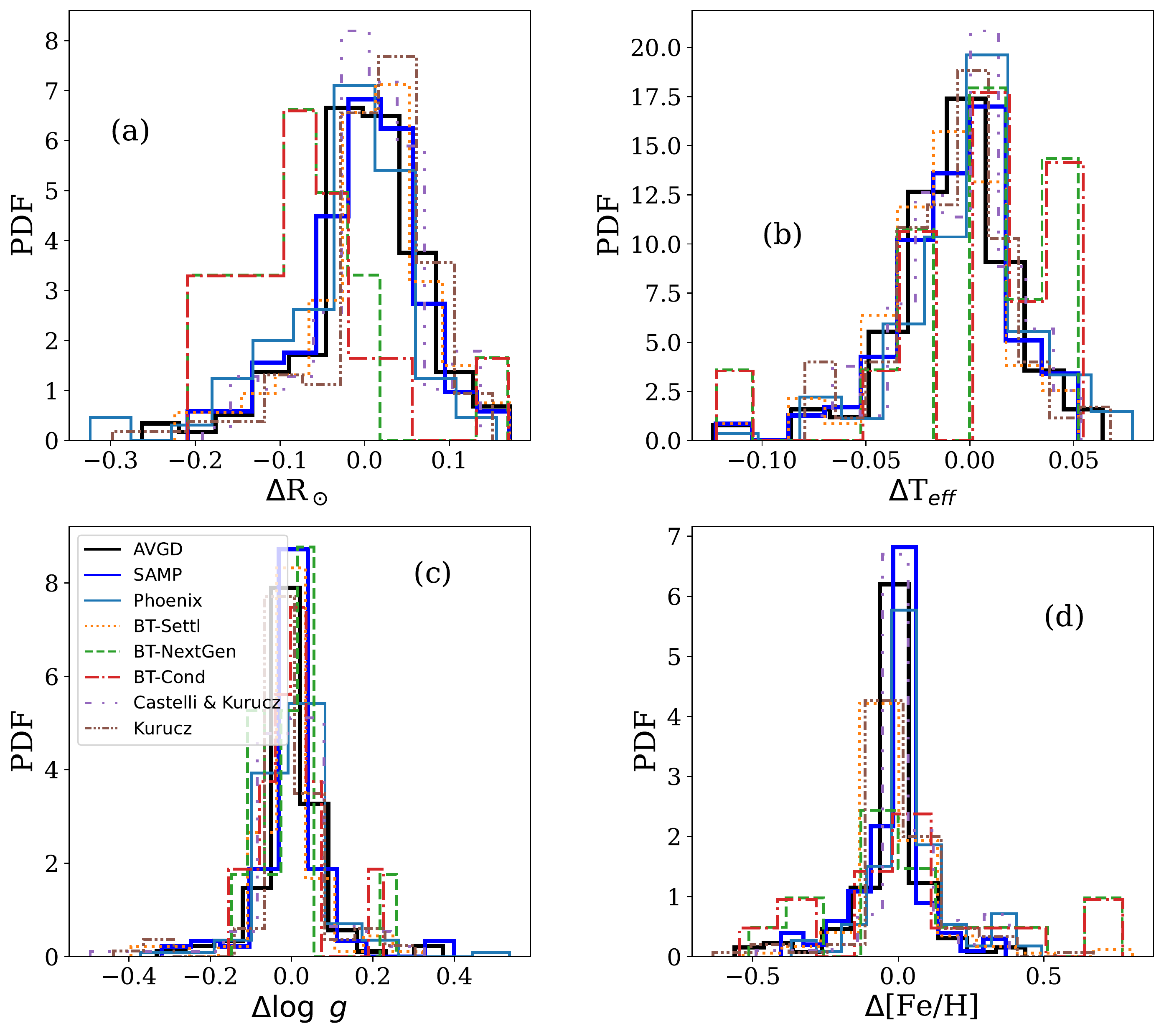}
    \caption{Offset distributions for each model for (a) stellar radius, (b) effective temperature, (c) $\log{\rm g}$ and (d) [Fe/H]. The averaged distribution is plotted to show how it decreases individual model biases, except in the case of giant stars where the models consistently overestimate radius and temperature.}
    \label{fig:distributions}
\end{figure*}

An important aspect of this work is highlighting the systematic offsets of the different models used, and to uncover possible biases present in them. To this end we studied the fractional residual distributions for the radius and temperature, and the residual distributions for the $\log{\rm g}$ and [Fe/H]. Ideally the residuals should have a normal distribution with a mean of zero, however, we show in Figure \ref{fig:distributions} that the residual distributions are slightly asymmetrical and skewed, likely highlighting biases in the models, but without a significant offset.
From Table \ref{tab:offsets} we see that none of the parameters show a significant offset in their residuals for both the full and truncated sample, but some of the individual models can over or underestimate measurements by as much as 32\% for radius and 12\% for temperature which translates to $\sim0.6$~R$_\odot$ and 542~K, 0.54 dex and 0.8 dex for $\log~g$ and [Fe/H] respectively.

\begin{table}
\caption{Individual model offsets for the sample of stars.}              
\label{tab:offsets}      
\centering                                     
\begin{tabular}{l c c c c c}          
\hline\hline          
Model & mean & spread & -3$\sigma$ & +3$\sigma$ & max\\
\hline 
& \multicolumn{5}{c}{Radius} \\ 
\hline
Phoenix v2 & -0.02 & 0.08 & 0.32 & 0.15 & 0.32 \\
BT-Settl & 0.01 & 0.07 & 0.22 & 0.17 & 0.22 \\
BT-NextGen & -0.07 & 0.09 & 0.21 & 0.17 & 0.21 \\
BT-Cond & -0.07 & 0.09 & 0.21 & 0.17 & 0.21 \\
Castelli \& Kurucz & 0.00 & 0.06 & 0.19 & 0.14 & 0.19 \\
Kurucz & 0.01 & 0.07 & 0.29 & 0.15 & 0.30 \\
\hline
& \multicolumn{5}{c}{Temperature} \\
\hline
Phoenix v2 & 0.00 & 0.03 & 0.11 & 0.08 & 0.12 \\
BT-Settl & -0.01 & 0.03 & 0.12 & 0.05 & 0.12 \\
BT-NextGen & 0.00 & 0.04 & 0.12 & 0.05 & 0.12 \\
BT-Cond & 0.00 & 0.04 & 0.12 & 0.05 & 0.12 \\
Castelli \& Kurucz & -0.00 & 0.03 & 0.08 & 0.05 & 0.08 \\
Kurucz & -0.01 & 0.03 & 0.08 & 0.07 & 0.08 \\
\hline
& \multicolumn{5}{c}{$\log{\rm g}$} \\ 
\hline
Phoenix v2 & 0.01 & 0.09 & 0.34 & 0.49 & 0.54 \\
BT-Settl & -0.01 & 0.09 & 0.38 & 0.31 & 0.39 \\
BT-NextGen & -0.00 & 0.10 & 0.15 & 0.25 & 0.26 \\
BT-Cond & -0.01 & 0.09 & 0.15 & 0.22 & 0.23 \\
Castelli \& Kurucz & 0.00 & 0.10 & 0.48 & 0.32 & 0.49 \\
Kurucz & -0.00 & 0.10 & 0.43 & 0.30 & 0.44 \\
\hline
& \multicolumn{5}{c}{[Fe/H]} \\ 
\hline
Phoenix v2 & 0.04 & 0.14 & 0.36 & 0.48 & 0.49 \\
BT-Settl & 0.01 & 0.15 & 0.50 & 0.79 & 0.80 \\
BT-NextGen & 0.06 & 0.35 & 0.51 & 0.77 & 0.77 \\
BT-Cond & 0.07 & 0.35 & 0.54 & 0.77 & 0.77 \\
Castelli \& Kurucz & 0.03 & 0.12 & 0.47 & 0.38 & 0.49 \\
Kurucz & 0.01 & 0.15 & 0.61 & 0.62 & 0.67 \\
\hline
\end{tabular}
\end{table}

The biases highlighted in Figure \ref{fig:distributions} arise from the different input physics and geometrical assumptions of each model, and while the BT-models have similar inputs and solar abundances, these subtle differences are particularly sensitive to cool, low-mass stars, likely a consequence of the grand differences between their internal structures when compared to that of the Sun\footnote{Many models are benchmarked against the Sun due to the deep understanding we have of the Solar structure.}.  A thorough review on current topics and shortcomings in stellar modeling can be found in Section 6.3 of \citet{Torres2010} and the discussions found in \citet{Jofre14, Jofre15, Heiter15, Jofre17}. 

The lack of a significant offset in the averaged residual distribution is evidence of the accuracy \texttt{ARIADNE} achieves with the radius, temperature and $\log{\rm g}$. The small spread found in the averaged residual distributions show the precision attainable by this method.  Indeed, some of the posterior widths our code attains is at the level of, or better than, those directly measured from interferometry.  Therefore, the widespread use of \texttt{ARIADNE} for obtaining stellar parameters will allow us to calculate more accurate and precise planetary masses and radii, further constraining planetary formation, evolution and composition models.

In \cite{Tayar2020}, while not studying stellar atmosphere grids directly, they investigate the uncertainties of stellar fundamental properties for FGK stars. They advocate for additional uncertainties added in quadrature to the formal uncertainties for stars, due in some part to model uncertainties. \texttt{ARIADNE} circumvents part of this problem, since we include this error as part of the modeling procedure within the BMA. What \citeauthor{Tayar2020} did not consider, is the constraints from the data themselves, since the precision clearly allows models to be favored over other models, in many cases. Therefore, by properly weighting each model given the constraints from the data, this systematic variability can be partially overcome. We note, however, that due to the inherent complexity of stellar atmosphere modeling, quantifying how much of this systematic variability is overcome is a difficult task.

\section{Conclusions}
\label{sec:conclusion}

We present \texttt{ARIADNE}, a public and open-source \texttt{Python} package to perform automatic stellar SED fitting using Bayesian Model Averaging to calculate stellar parameters with precision and accuracy comparable to interferometric observations.  We find a fractional RMS of $0.001\pm0.070~$ and a mean precision of $2.1\%$ for the averaging method and $0.000\pm0.069~$ and $4.3\%$ for the sampling method when comparing the results of our set of 135 stars to those measured by interferometry. Considering an expected scatter in interferometric precision of 5\% \citep{Stassun2017}, we conclude that the scatter found when comparing \texttt{ARIADNE} with interferometric measurements can be mostly explained by the intrinsic noise within these measurements. The code is particularly useful for calculating stellar radius, effective temperature with uncertainties lower than $\pm$100~K, and $\log{\rm g}$, which can then be used to calculate mass either through Equation \ref{eq:grav_mass_log} or using isochrone interpolation. Accurate metallicity values can be attained when using priors drawn from other methods, such as EW calculations or spectral synthesis, though these values might not be as precise as when using these other methods. These are all key parameters for characterizing planetary bulk parameters in the age of high precision and short-cadence ground- and space-based photometry.

We highlighted systematic differences between the employed models, and found that while the different assumed input physics and geometries give rise to non-negligible biases, these biases are mostly overcome when properly averaging the models with their respective Bayesian evidences. This does not apply to the isochrone masses \texttt{ARIADNE} provides, but the same methodology can be applied to different isochrone models to increase the precision of estimated masses and ages.  The M dwarfs seem to  pose something of a problem also, showing a rising trend in radii when compared to interferometrically measured values, but the effect is small, well within the 5\% interferometric scatter.  We finally encourage the use of \texttt{ARIADNE} to calculate stellar bulk properties, along with continued development of the methodologies presented here within the framework of stellar and exoplanetary science.

\section*{Acknowledgements}
JIV acknowledges support of CONICYT-PFCHA/Doctorado Nacional-21191829.
JSJ acknowledges support by FONDECYT grant 1201371 and partial support from the ANID Basal project FB210003.  This research has made use of the VizieR catalogue access tool, CDS, Strasbourg, France (DOI: 10.26093/cds/vizier). The original description of the VizieR service was published in A\&AS 143, 23.

\section*{Data Availability}
The data underlying this article are available in the article and in its online supplementary material.




\bibliographystyle{mnras}
\bibliography{ref} 






\bsp	
\label{lastpage}
\end{document}